\documentclass[12pt,preprint]{aastex}

\newcommand{\km}{{\rm\,km}}

\newcommand{\rad}{\rm\,rad}
\newcommand{\AU}{{\rm\, AU}}

\newcommand{\yr}{{\rm\,yr}}
\newcommand{\YR}{{\rm yr}}

\newcommand{\pomega}{\tilde{\omega}}

\begin{document}

\shortauthors{Chiang and Jordan}
\shorttitle{Plutinos and Twotinos}

\title{On the Plutinos and Twotinos of the Kuiper Belt}

\author{E.~I.~Chiang \& A.~B.~Jordan}

\affil{Center for Integrative Planetary Sciences\\
Astronomy Department\\
University of California at Berkeley\\
Berkeley, CA~94720, USA}

\email{echiang@astron.berkeley.edu, abjordan@uclink.berkeley.edu}

\begin{abstract}
We illuminate dynamical properties of
Kuiper Belt Objects (KBOs) in the 3:2 (``Plutino'')
and 2:1 (``Twotino'') Neptunian resonances within the model
of resonant capture and migration. We analyze a
series of numerical integrations,
each involving the 4 migratory giant planets and 400 test particles
distributed throughout trans-Neptunian space,
to measure the efficiencies of
capture into the 3:2 and 2:1 resonances,
the efficiencies of capture
into Kozai-type secular resonances, and the libration centers
and amplitudes of resonant particles, all as functions
of the migration speed. We synthesize instantaneous snapshots
of the spatial distribution of $\sim$$10^4$ resonant KBOs, from
which we derive the longitudinal variation of the sky density
of each resonant family. Twotinos cluster $\pm 75\degr$ away
from Neptune's longitude, while Plutinos cluster $\pm 90\degr$ away.
Such longitudinal clustering
persists even for surveys that are not volume-limited
in their ability to detect resonant KBOs.
Remarkably, between $-90\degr$ and $-60\degr$ of Neptune's longitude, we
find the ratio of sky densities of Twotinos to Plutinos
to be nearly unity despite
the greater average distance of Twotinos, assuming the two
resonant populations are equal in number and share the same
size, albedo, and inclination distributions.
We couple our findings to observations
to crudely estimate that the intrinsic Twotino population
is within a factor of $\sim$3 of the Plutino population.
Most strikingly, the migration model predicts a possible asymmetry in
the spatial distribution of Twotinos: more Twotinos
may lie at longitudes behind that of Neptune than
ahead of it. The magnitude of the asymmetry amplifies dramatically
with faster rates of migration and can be as large as $\sim$300\%.
A differential measurement of the sky density of 2:1 resonant objects
behind of and in front of Neptune's longitude would powerfully
constrain the migration history of that planet.
\end{abstract}


\section{INTRODUCTION}
\label{intro}

The substantial eccentricity, $e_p$,
of Pluto's orbit is well explained by Malhotra's (1995) theory
of resonant capture by Neptune. In this scenario, Neptune
migrated radially outwards from the sun by scattering planetesimals
towards Jupiter, captured Pluto into its 3:2 mean-motion resonance,
and amplified $e_p$ upon continuing its migration.
The resonant amplification of $e_p$
can be understood either mechanistically using Gauss's equations
(see, e.g., Peale 1986), or in terms of the preservation of an
adiabatic invariant (see, e.g., Yu \& Tremaine 2001).
The discovery of dozens of
Kuiper Belt Objects (KBOs) that share the 3:2 Neptunian
resonance with Pluto and that also exhibit large orbital
eccentricities (Jewitt \& Luu 2000)
apparently vindicates this proposal that Neptune
plowed its way outwards through a field of planetesimals
early in the history of the solar system (Fernandez \& Ip 1984).

In the particular numerical simulation presented by Malhotra (1995),
the 2:1 Neptunian resonance is predicted to be about as equally populated
as the 3:2 resonance. For brevity, we will refer
to KBOs in the latter resonance as ``Plutinos''
and KBOs in the former resonance as ``Twotinos.''
As of July 4, 2002, the Minor Planet
Center (MPC) database contains $\sim$41 Plutino
candidates (objects observed at multiple
oppositions having fitted semi-major axes, $a$,
within 0.3 AU of the exact 3:2 resonance
location at $a_{3:2} = 39.5 \AU$), and $\sim$5 Twotino candidates
(objects whose $a$'s lie within 0.3 AU of the exact 2:1 resonance location
at $a_{2:1} = 47.8
\AU$).\footnote{\url{http://cfa-www.harvard.edu/iau/lists/TNOs.html}}
The Twotino candidates possess substantial orbital eccentricities,
$e \approx 0.2$--$0.4$, in accord with the predictions of
resonant capture and migration.
Membership in a resonance is confirmed by verifying that the
appropriate resonant argument
librates rather than circulates (see \S\ref{gencon}); an orbit
classification scheme based on this more rigorous
criterion is currently being developed by the Deep Ecliptic
Survey (DES) Team (see Millis et al.~2002).
For the present paper, we will consider the
{\it observed} Twotinos to be outnumbered by the
{\it observed} Plutinos by a factor of $F_{obs} \sim 8$.
Part of this bias must simply reflect the
fact that $a_{2:1} > a_{3:2}$; all other factors
being equal, more distant objects
are fainter and more difficult to detect. But
part of this bias may also reflect selection
effects that depend on the longitude
and latitude of observation. A mean-motion resonant object
will be preferentially found at certain locations
with respect to Neptune---what we will call
``sweet spots'' on the sky. The sweet spots for Twotinos
are not necessarily those of Plutinos.

Ida et al.~(2000) point out that Neptune's ability
to resonantly capture objects
varies with migration timescale. A migration timescale
that is 20 times shorter than that considered
by Malhotra (1995) is found to severely reduce
the probability of capture into the 2:1 resonance.
For the 3:2 resonance, the capture probability
is affected less dramatically by reductions in
the migration timescale. The relative robustness
of the 3:2 resonance compared to the 2:1 resonance is explored
analytically by Friedland (2001), who underscores
the importance of the indirect potential for the
latter resonance.

This paper quantifies, within the confines of the model
of resonant migration, the bias against finding KBOs
in the 2:1 resonance over those in the 3:2 resonance.
In \S\ref{gencon}, we set forth general, model-independent
considerations for calculating this bias. In \S\ref{migmod},
we describe in detail the results of a particular simulation
of resonant migration. In this section, we present illustrative
snapshots of the instantaneous spatial distributions of
Twotinos and of Plutinos. In \S\ref{vary}, we explore
how our results change by varying the migration rate of Neptune.
In \S\ref{conc}, we discuss our theoretical results in the
context of the observations. There, we begin to examine critically
the belief that Plutinos intrinsically outnumber Twotinos.
A summary of our main findings is provided in \S\ref{summa}.
Our computations may serve not only to de-bias extant observations
and thereby constrain the true relative resonant populations,
but also to guide future observational surveys.

\section{GENERAL CONSIDERATIONS}
\label{gencon}

The ability of an observational survey to detect KBOs residing
within a given resonance depends on the KBOs' (1) spatial distribution,
and (2) size and albedo distributions. In this
section, we offer comments regarding the former consideration.

\subsection{Mean-Motion Resonances}
\label{mmr}

By definition, an object inhabits a $j\!+\!1$:$j$ outer Neptunian
resonance if the resonant argument,

\begin{equation}
\Phi_{j+1:j} \equiv (j+1)\lambda - j\lambda_N - \pomega \, ,
\end{equation}

\noindent librates (undergoes a bounded oscillation about a particular
angle). Here $j$ is a positive integer, $\lambda$ and
$\pomega$ are the mean longitude and longitude of periastron
of the object, respectively, and $\lambda_N$ is the mean longitude
of Neptune. A restricted range for $\Phi_{j+1:j}$ implies
that the resonant particle will most likely be found at particular
longitudes with respect to Neptune. For example, if an
object inhabits the $j=1$ resonance such that $\Phi_{2:1}$
librates about $180\degr$ with a negligibly small libration
amplitude, then such an object
attains perihelion when Neptune is $180\degr$ away in longitude.
The eccentricity of the resonant object may be so large that
the orbits of Neptune and of the particle cross, but the particle
avoids close encounters with Neptune by virtue of the boundedness
of $\Phi_{2:1}$.

\begin{figure}
\epsscale{0.6}
\plotone{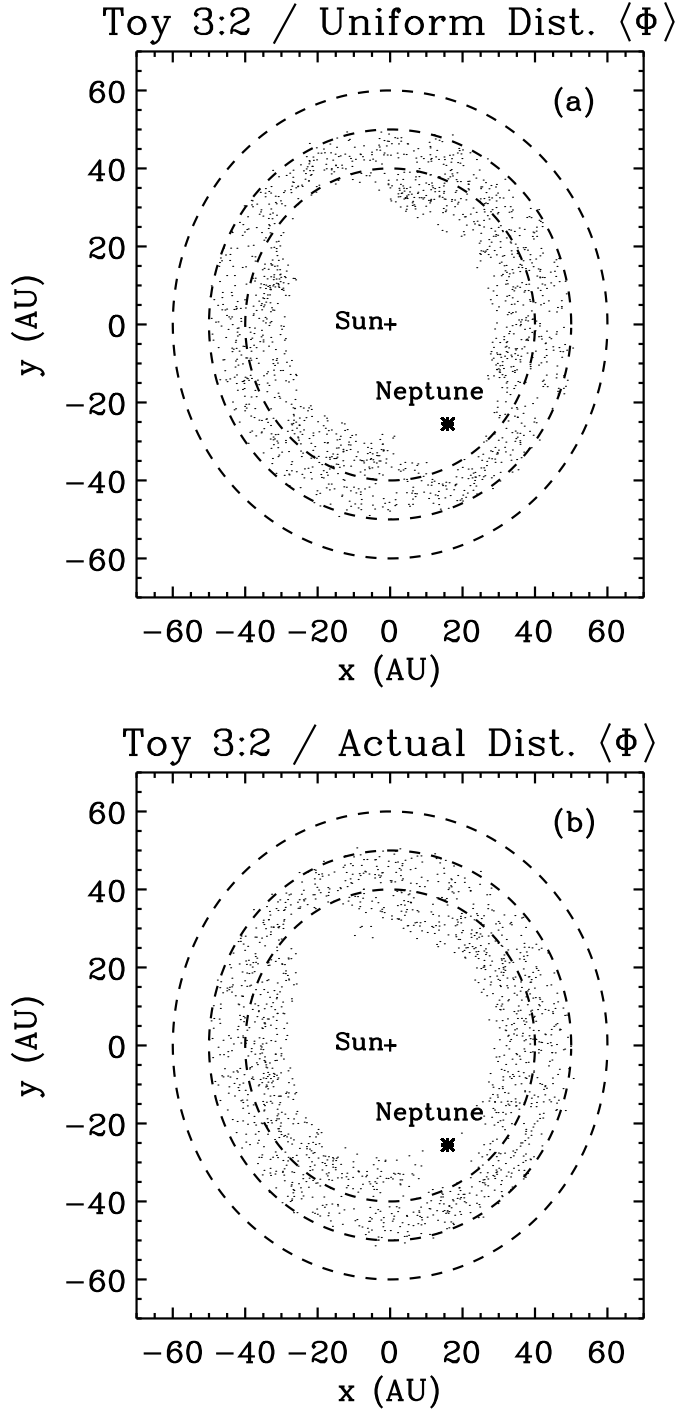}
\caption{Toy models for the spatial distribution of Plutinos.
The particles' resonant arguments equal $\Phi_{3:2} = \pm 180\degr
+ \Delta \Phi_{3:2} \sin A$, where $A$ is uniformly distributed between
$0\degr$
and $360\degr$. In panel (a), $\Delta\Phi_{3:2}$ is
uniformly distributed between $100\degr$ and $120\degr$.
In panel (b), $\Delta\Phi_{3:2}$ reflects the distribution
obtained through simulation Ia, as shown by the solid histogram of
Figure \ref{histo32allin1}. Where Plutinos cluster depends sensitively on
the distribution of $\Delta\Phi_{3:2}$. The dashed circles delimit
radii of 40, 50, and 60 AU.
\label{toy32}}
\end{figure}

It has been remarked that because $\Phi_{3:2}$ for Pluto
and the Plutinos librates about a mean value of
$\langle \Phi_{3:2} \rangle = 180\degr$, these objects
tend to be found at longitudes
displaced $\pm 90\degr$ from Neptune when they reach perihelion
and are at their brightest (e.g., Jewitt, Luu, \& Trujillo 1998).
This argument is not strictly correct; it neglects
the Plutinos' often substantial libration amplitudes,
$\Delta \Phi_{3:2}$. Just as a librating pendulum is most
likely found near the turning points of its trajectory,
a Plutino's resonant argument is most likely found
near $\langle \Phi_{3:2} \rangle + \Delta \Phi_{3:2}$
or near $\langle \Phi_{3:2} \rangle - \Delta \Phi_{3:2}$,
not $\langle \Phi_{3:2} \rangle$.
Figures \ref{toy32}a and \ref{toy32}b portray two
toy models for the spatial distributions of 3:2 resonant objects.
They demonstrate that the spatial distribution of resonant particles is
sensitive to the distribution of libration amplitudes,
$dN/d\Delta \Phi$, and not just to the value of the libration center,
$\langle \Phi \rangle$. For each panel, the instantaneous locations
of 15000 co-planar particles are calculated according to the
following scheme: semi-major axes are randomly chosen from
a uniform distribution between 39.0 and 39.8 AU, mean longitudes
are randomly drawn from a uniform distribution between $0\degr$ and
$360\degr$, and eccentricities are randomly selected from a uniform
distribution between 0.1 and 0.3. Resonant arguments of particles
are taken to equal $\Phi_{3:2} = \pm 180\degr + \Delta \Phi_{3:2} \sin A$,
where the
upper and lower signs are equally probable and $A$ is randomly
drawn from a uniform distribution between $0\degr$ and $360\degr$.
For Figure \ref{toy32}a, $\Delta \Phi_{3:2}$ is randomly
selected from a uniform distribution between $100\degr$ and
$120\degr$, and for Figure \ref{toy32}b, the underlying
distribution for $\Delta \Phi_{3:2}$ is given by the solid
histogram in Figure \ref{histo32allin1}.
The longitude of perihelion of each particle is calculated
according to $\pomega = 3\lambda - 2\lambda_N - \Phi_{3:2}$, where
$\lambda_N$ is assigned its present-day value of 302$\degr$.

\placefigure{fig1}

The resultant plots illustrate two ways that Plutinos
could be distributed, both of which are possible in principle.
In Figure \ref{toy32}a, the objects cluster in 4 locations,
respecting the $2\times 2$ turning points of the resonant
argument---2 turning points for each of the 2 libration
centers, $\langle \Phi_{3:2} \rangle = \pm 180\degr$. Thus, it
is not true {\it a priori} that 3:2 resonant objects
must cluster at only two locations in the sky.
By contrast,  in Figure \ref{toy32}b,
there are enough small-amplitude librators $(\Delta \Phi_{3:2} < 1 \rad)$
that the concentration of objects does gently peak at longitudes
$\pm 90\degr$ away from Neptune, in abeyance with the usual expectation.

\begin{figure}
\epsscale{1.1}
\plotone{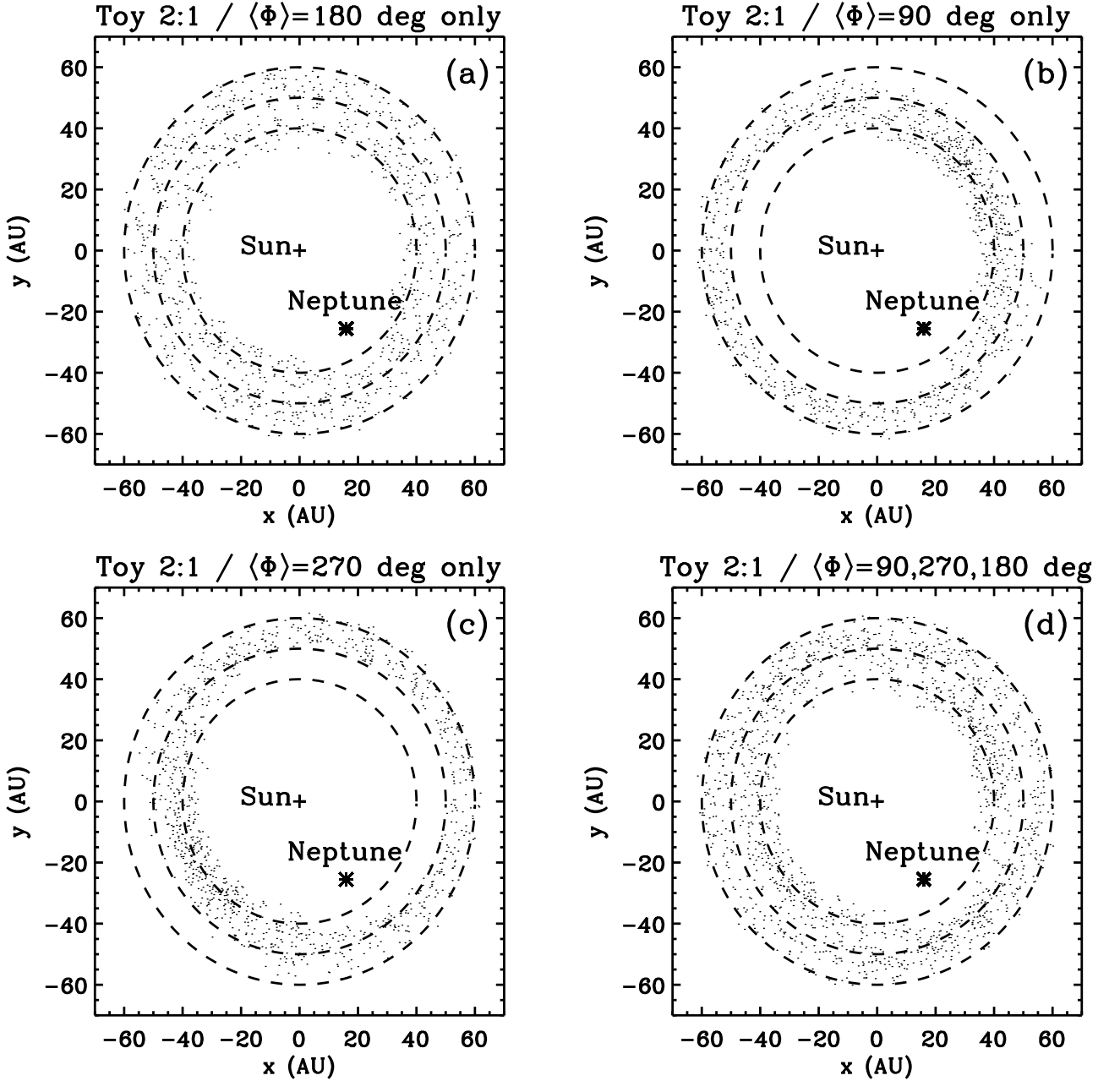}
\caption{Toy models for the spatial distribution of Twotinos.
Each panel portrays particles having libration centers,
$\langle \Phi_{2:1} \rangle$, that are indicated by the panel title.
For $\langle \Phi_{2:1} \rangle = 180\degr$, particles avoid
Neptune's longitude and are distributed symmetrically about the
Sun-Neptune line.
For $\langle \Phi_{2:1} \rangle = 90\degr \, (270\degr)$,
particles cluster $90\degr$ ahead (behind) of Neptune's
longitude. The combination of these libration centers in the proportions
found in simulation Ib yields panel (d),
which resembles Figure \ref{snap21}.
Toy model parameters are uniformly distributed over
the following ranges: $47.2 \AU \leq a \leq 48.4 \AU$,
$0.1 \leq e \leq 0.3$, $0\degr \leq \lambda < 360\degr$,
$0\degr \leq A < 360\degr$, $135\degr \leq \Delta\Phi_{2:1} \leq 150\degr$
if $\langle \Phi_{2:1} \rangle = 180\degr$, and $15\degr \leq \Delta\Phi_{2:1}
\leq 60\degr$ otherwise.
\label{toy21}}
\end{figure}

Toy models such as these are useful for analyzing the results of numerical
orbit integrations. In anticipation of such integrations, we present
Figure \ref{toy21}, which displays a toy model for the distribution
of Twotinos. The parameters of the model are described in the
figure caption.

\placefigure{fig2}

\subsection{Secular Resonances}
\label{secres}

Secular resonances might also play a role in determining
the spatial distribution of KBOs. Chief among these are
Kozai-type resonances in which $\omega$, the argument of
perihelion, librates about particular angles, usually
$\pm 90\degr$, $0\degr$, or $180\degr$. Pluto inhabits
a Kozai resonance established by the total secular potential
of all 4 giant planets, such that
its $\omega$ librates about 90$\degr$ with an amplitude
of $23\degr$ (for a review, see Malhotra \& Williams 1997).
Thus, Pluto attains perihelion and is brightest
only when it sits above the invariable plane by
its orbital inclination of $i_p \approx 16\degr$.
If enough Plutinos inhabit Pluto-like Kozai resonances,
their detection would be influenced by
selection effects that depend on the latitude of observation.

How many Plutinos and Twotinos inhabit Kozai-type resonances?
Nesvorny, Roig, \& Ferraz-Mello (2000) find that
few observed Plutinos, $\sim$2--4 out of 33,
exhibit libration of $\omega$. They explore a scenario
by which Pluto gravitationally scatters other Plutinos
out of the Kozai resonance. They find the scenario to be
viable, though whether it is required by the model
of resonant capture and migration is unknown;
the efficiency of capture into a Pluto-like Kozai resonance
in Malhotra's (1995) model of planetary migration might
already be small enough to explain the observations.
Unlike the case for the 3:2 resonance,
we are not aware of any study of
the possibility of $\omega$-libration within the 2:1 resonance.
In \S\ref{secresres} below,
we investigate by direct numerical simulation
the capture probability into Kozai-type resonances
for both Plutinos and Twotinos within the model of planetary migration.

\section{MIGRATION MODEL}
\label{migmod}

Here we describe our model
for the radial migration of the four giant planets
and the resonant capture of planetesimals.
Model ingredients are supplied in \S\ref{ingred},
mean-motion resonance capture efficiencies are computed
in \S\ref{capeff}, mean-motion resonance libration statistics and retainment
efficiencies are discussed in \S\ref{libstat},
and the statistics of secular resonance capture are
presented in \S\ref{secresres}.
Those readers interested in the spatial distribution of resonant
objects may skip to \S\ref{snap} without much loss of continuity.

\subsection{Initial Conditions and Migration Prescription}
\label{ingred}
To effect the migration, we follow Malhotra (1995) and introduce
a perturbative acceleration on each planet of the form

\begin{equation}
\delta \ddot{r} = - \frac{\hat{v}}{\tau} \left( \sqrt{\frac{GM_{\odot}}{a_f}} -
\sqrt{\frac{GM_{\odot}}{a_i}} \right) \exp{(-t/\tau)} \, ,
\label{migacc}
\end{equation}

\noindent where $a_i$ and $a_f$ are the initial and final semi-major
axes of a given planet, respectively, $G$ is the gravitational constant,
$t$ measures time, $\tau$ is a time constant, and $\hat{v}$ is the unit
vector pointing in the instantaneous direction of the planet's velocity.
Equation (\ref{migacc}) corrects a typographical sign error in equation (7) of
Malhotra (1995). This prescription causes each planet's semi-major
axis to evolve according to

\begin{equation}
a(t) = a_f - (a_f - a_i) \exp{(-t/\tau)} \, ,
\end{equation}

\noindent but does not directly induce long-term changes in the
planet's eccentricity and inclination. We adopt values for
$(a_i,a_f)$ for each of the planets as follows (in AUs): Jupiter
$(5.00,5.20)$, Saturn $(8.78,9.58)$, Uranus $(16.2,19.2)$,
and Neptune $(23.1,30.1)$.

We work in a coordinate system that takes the reference
plane to be the invariable plane of the solar system.
The positions and velocities of each planet
at $t=0$ are adapted from Cohen, Hubbard, and Oesterwinter (1973),
with the positions multiplied by $a_i/a_f$ and the velocities
multiplied by $\sqrt{a_f/a_i}$. We employ the symplectic
integrator, SyMBA, developed by Duncan, Levison, \& Lee (1998),
which is based on the algorithm by Wisdom \& Holman (1991).
The integrator was kindly supplied to us by E. Thommes,
M. Duncan, \& H. Levison.

For simulations Ia and Ib that are described in \S\ref{migmod},
we take $\tau = 10^7 \yr$. Shorter migration periods
of $\tau = 10^6 \yr$ and $\tau = 10^5 \yr$ are considered
in \S\ref{vary}.

In simulation Ia, we focus on the efficiency of capture into,
and the resultant dynamics within, the 3:2 resonance. We introduce
400 massless test particles whose initial semi-major axes lie between
31.4 AU ($=1\AU$ greater than the initial location of the 3:2 resonance)
and 38.5 AU ($=1\AU$ short of the final location of the 3:2 resonance).
All particles in this region have the potential to be captured
into the sweeping 3:2 resonance. Their initial eccentricities and inclinations
are randomly drawn from uniform distributions between 0 and 0.05, and
between $0\degr$
and $1.4\degr = 0.025 \rad$, respectively. Arguments of periastron
($\omega$), longitudes of ascending nodes ($\Omega$), and mean
anomalies ($M$) are uniformly sampled between 0 and $2\pi$. The duration
of the integration spans $t_f^{\rm{Ia}} = 6 \times 10^7 \yr = 6\tau$.

In simulation Ib, we concentrate on the 2:1 resonance. The only
essential difference between simulations Ia and Ib is that for the latter,
the 400 test particles are distributed initially between
37.7 AU ($=1\AU$ greater than the initial location of the 2:1 resonance)
and 46.8 AU ($=1\AU$ less than the final location of the 2:1 resonance).
Thus, all such particles are potentially captured into the 2:1
resonance. The duration of this simulation
is $t_f^{\rm{Ib}} = 8 \times 10^7 \yr = 8 \tau$.

\subsection{Capture Efficiencies}
\label{capeff}

\begin{figure}
\epsscale{0.8}
\plotone{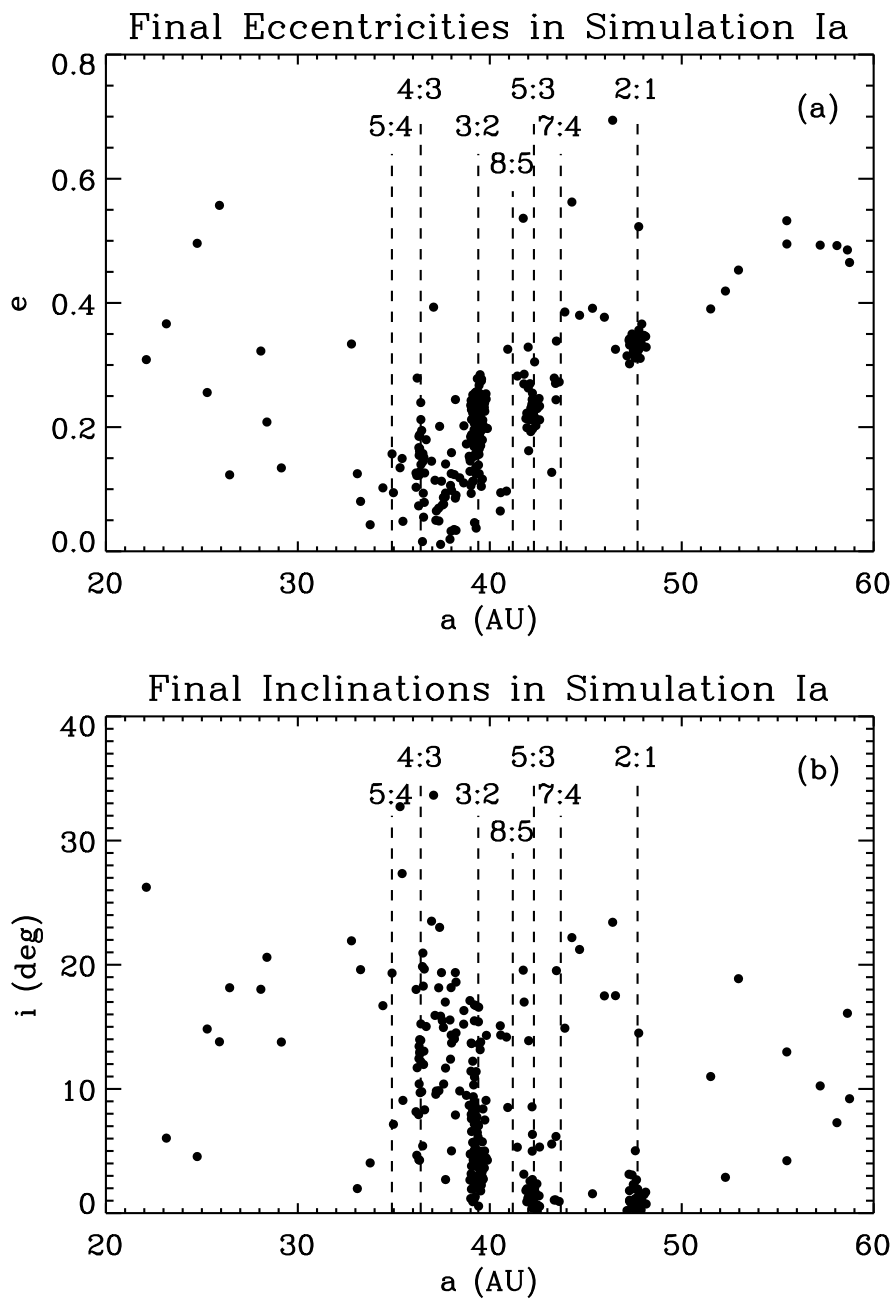}
\caption{(a) Final eccentricities vs.~semi-major axes
for particles with $a \leq 60 \AU$
in simulation Ia for which $\tau = 10^7\yr$.
(b) Final inclinations vs.~semi-major axes. Out of
400 test particles potentially swept into the
3:2 resonance, 92 are actually captured. Of these
92, perhaps only 42 would remain bound to the 3:2 resonance
over the age of the solar system.
\label{ea32}}
\end{figure}

Of the 400 test particles in simulation Ia, 92 are captured
into the 3:2 resonance. By definition,
$\Phi_{3:2}$ librates for these 92 objects but circulates
for the remaining 308. This capture efficiency of
$f_{3:2} \approx 23\%$ reflects (1) the probability
of capture into the isolated 3:2 resonant potential
just prior to resonance encounter (Henrard \& Lemaitre 1983,
Borderies \& Goldreich 1984),
(2) losses due to pre-emptive capture into ``competing''
resonances such as the 5:3 and 2:1 which lie exterior
to the 3:2, and (3) losses
due to violent scattering by close encounters with
the planets. Figure \ref{ea32} displays the final eccentricities
and inclinations versus the semi-major axes of those test
particles having $a \leq 60 \AU$.

\placefigure{fig3}

\begin{figure}
\epsscale{0.8}
\plotone{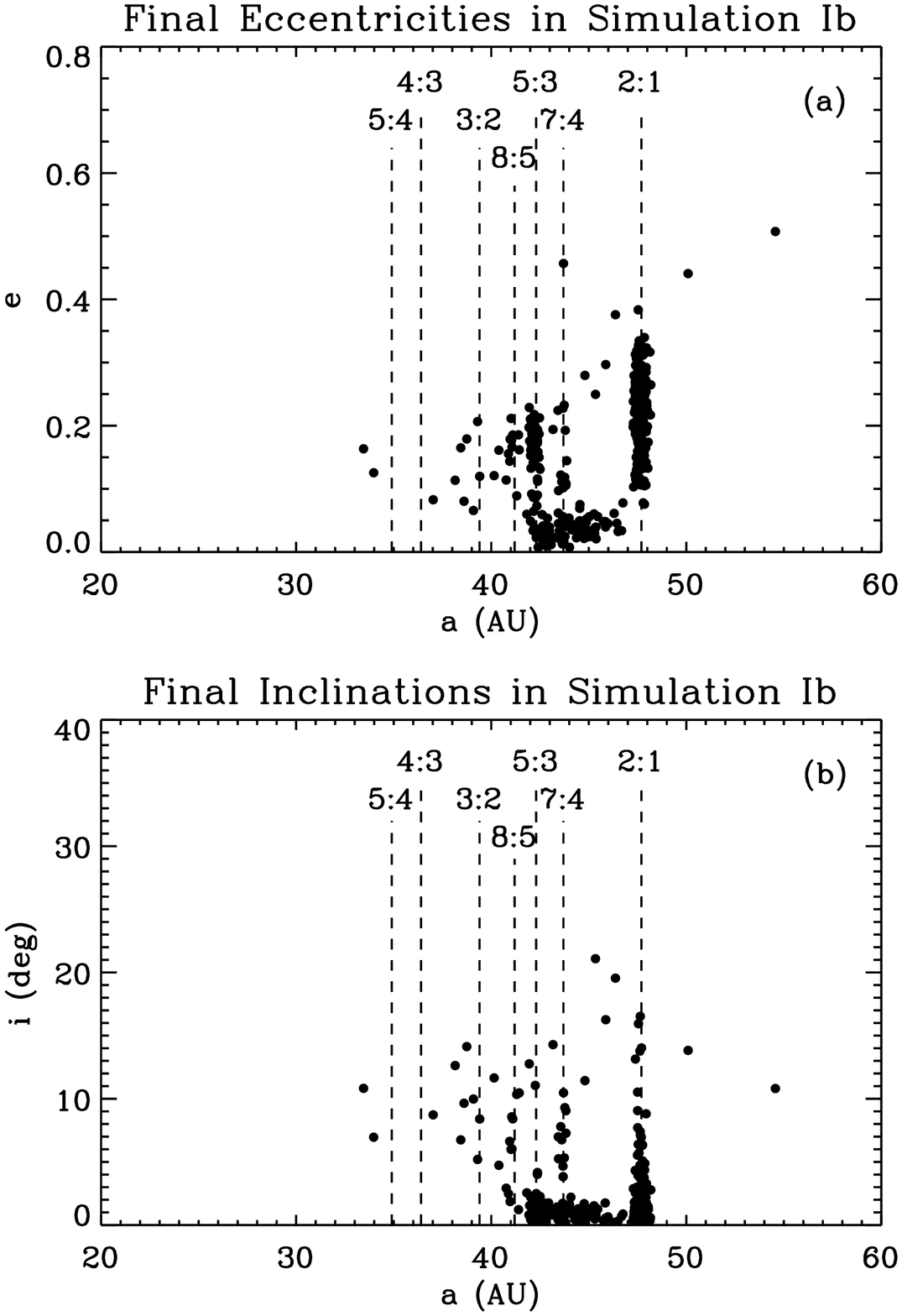}
\caption{(a) Final eccentricities vs.~semi-major axes
for particles with $a \leq 60 \AU$
in simulation Ib for which $\tau = 10^7\yr$.
(b) Final inclinations vs.~semi-major axes.
Out of 400 test particles potentially swept into the
2:1 resonance, 212 are actually captured. Of these
212, perhaps $\sim$100 would remain bound to the 2:1 resonance
over the age of the solar system.
\label{ea21}}
\end{figure}

{}From simulation Ib, we estimate the capture efficiency
of the 2:1 resonance to be $f_{2:1} \approx 212/400 = 53$\%.
This value is more than twice as high as $f_{3:2}$,
reflecting both the lack of competition from
other sweeping resonances which lie interior
to the 2:1, and the lower probability of scattering
by Neptune at these greater distances. Figure \ref{ea21}
displays the final ($a$,$e$,$i$) for those test particles
having $a \leq 60\AU$ at the close of simulation Ib.
Note that many objects remain uncaptured by any low-order
resonance at semi-major axes $43 \AU \lesssim a \lesssim 47 \AU$;
these non-resonant bodies presumably represent members
of the low-inclination Classical Kuiper Belt that is observed today
(Levison \& Stern 2001; Brown 2001).

\placefigure{fig4}

These capture efficiencies are recorded in Table \ref{capefftab}.
We emphasize that $f$ represents only the efficiency
of capture, as distinct from the efficiency of retainment
of captured objects, $g$, over the age of the solar
system. Whether a captured KBO remains in a given
resonance over $4\times 10^9\yr$ is discussed
in \S\ref{libstat}.

\begin{figure}
\epsscale{0.8}
\plotone{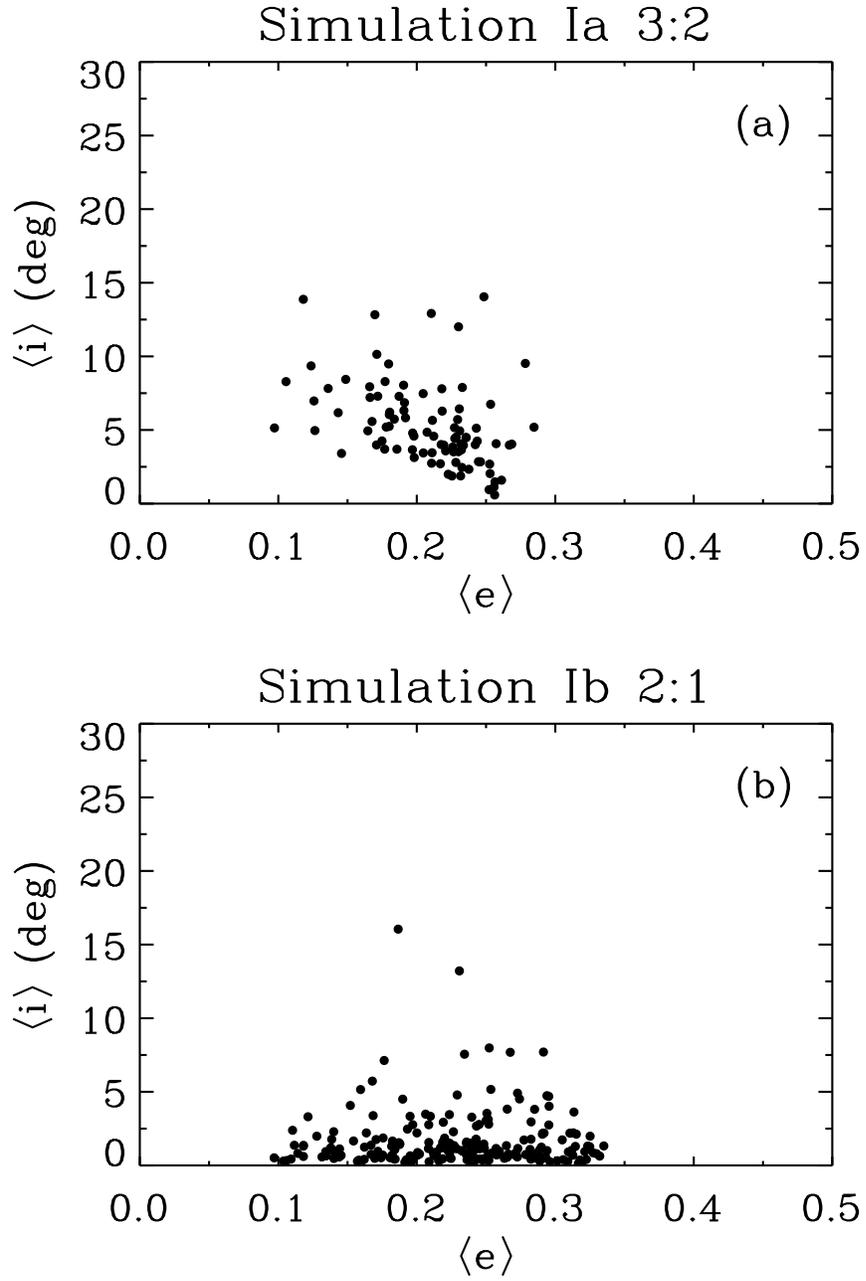}
\caption{(a) Average $i$ vs.~average $e$ of Plutinos during the final
$1 \times 10^7 \yr$ of simulation Ia, after the planets effectively
cease to migrate. (b) Average $i$ vs.~average
$e$ of Twotinos during the final $3 \times 10^7 \yr$
of simulation Ib. While $\langle i \rangle$ and $\langle e \rangle$
for Plutinos tend to be inversely correlated, no such tendency
exists for Twotinos.
\label{ei}}
\end{figure}

Mean inclinations, $\langle i \rangle$, of Plutinos and Twotinos
are plotted against mean eccentricities, $\langle e \rangle$,
in Figure \ref{ei}. The mean is taken over the last
$1\times 10^7\yr$ in simulation Ia, and the last $3\times 10^7\yr$
in Ib, during which times the migration has effectively stopped.
There is a tendency for the Plutinos to have their
$\langle i \rangle$'s and $\langle e \rangle$'s inversely
correlated. No apparent correlation exists between $\langle i \rangle$
and $\langle e \rangle$ for the Twotinos.
The predicted inclinations seem too low
to compare favorably with the observed
inclinations; the inadequacy of the migration
model in explaining the inclination distribution of Plutinos
has been noted by Brown (2001). We note further
that 4 out of the 5 observed Twotino candidates have orbital inclinations
between $11.8\degr$ and $13.5\degr$---values
characteristically larger than what the migration model predicts
for objects in the 2:1 resonance.

\placefigure{fig5}

\subsection{Libration Statistics and Retainment Efficiencies}
\label{libstat}

\begin{figure}
\epsscale{0.8}
\plotone{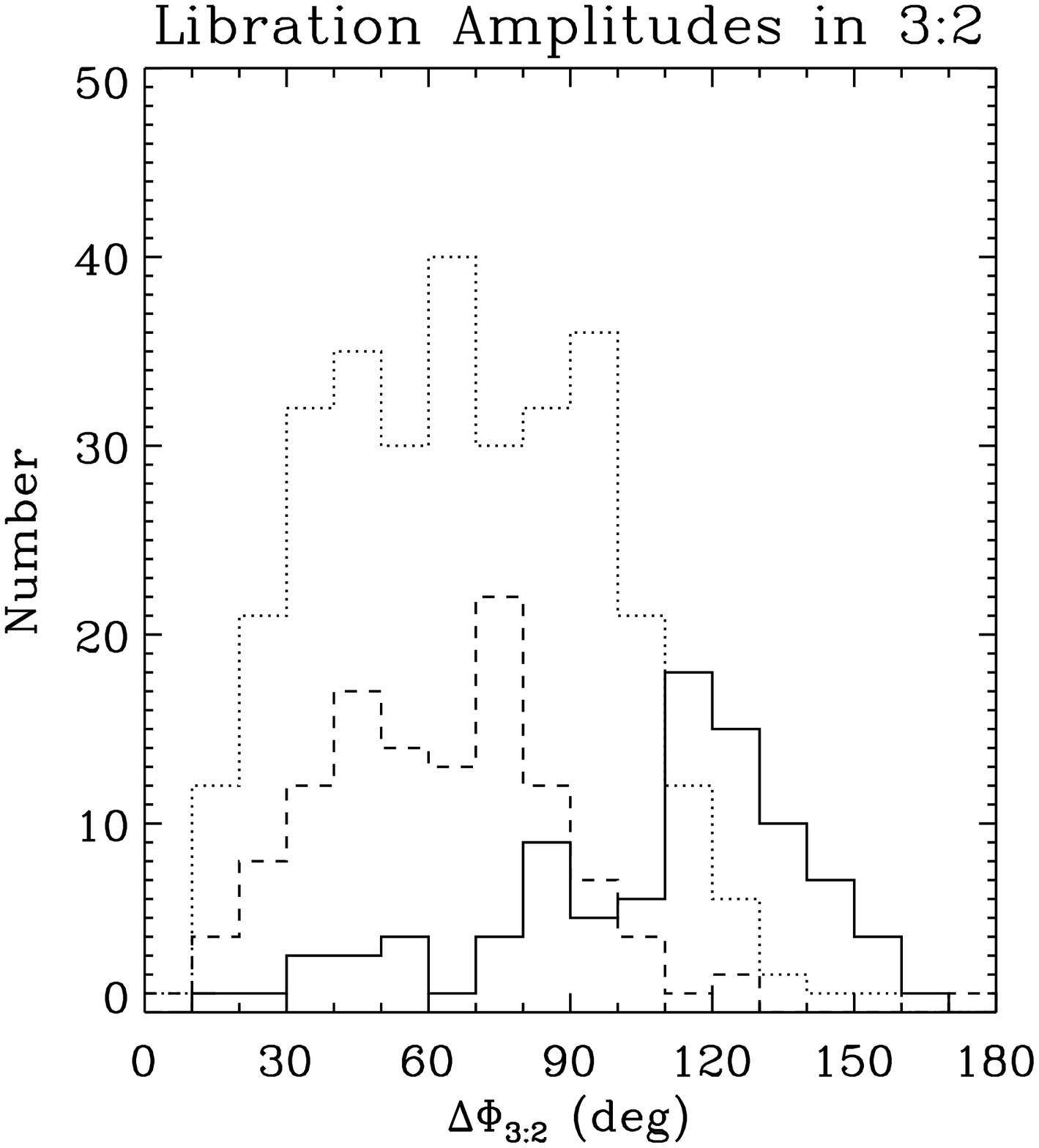}
\caption{Distribution of libration amplitudes of captured Plutinos in
simulation Ia (solid; $\tau = 10^7\yr$), IIa (dotted; $\tau = 10^6\yr$),
and IIIa (dashed; $\tau = 10^5\yr$). Simulations IIa and
IIIa are discussed in \S\ref{vary}. Reducing $\tau$ below
$10^7\yr$ yields smaller libration amplitudes and thus
greater efficiencies of retainment of
objects within the 3:2 resonance.
\label{histo32allin1}}
\end{figure}

For the Plutinos in simulation Ia, we find that
$\langle \Phi_{3:2} \rangle = \pi$, as expected from resonant
perturbation theory for low eccentricity orbits. The distribution
of libration amplitudes, $\Delta \Phi_{3:2} \equiv [\max (\Phi_{3:2}) -
\min (\Phi_{3:2})]/2$, is supplied in Figure \ref{histo32allin1}.
Most objects have substantial libration amplitudes $\gtrsim 1 \rad$.
Levison \& Stern (1995) and Morbidelli (1997) calculate that Plutinos
having large $\Delta \Phi$ can escape the 3:2 resonance over the
age of the solar system. Thus, many of the Plutinos that are
captured in our simulation Ia would not likely survive
if we were to extend our integration to $t_f^{\rm{Ia}} = 4\times 10^9\yr$.
We define the retainment efficiency, $g_{3:2}$,
to be the fraction of captured Plutinos that either
have $\Delta \Phi < 110\degr$ or that exhibit libration of
$\omega$ about $\pm 90\degr$. These selection criteria are
motivated by the stability study of Levison \& Duncan (1995; see
their Figure 7). Of the 92 captured Plutinos in simulation Ia,
42 satisfy our criteria for long-term residency; the resultant
value for $g_{3:2} = 46$\% is recorded in Table \ref{capefftab}.

\placefigure{fig6}

\begin{figure}
\epsscale{0.8}
\plotone{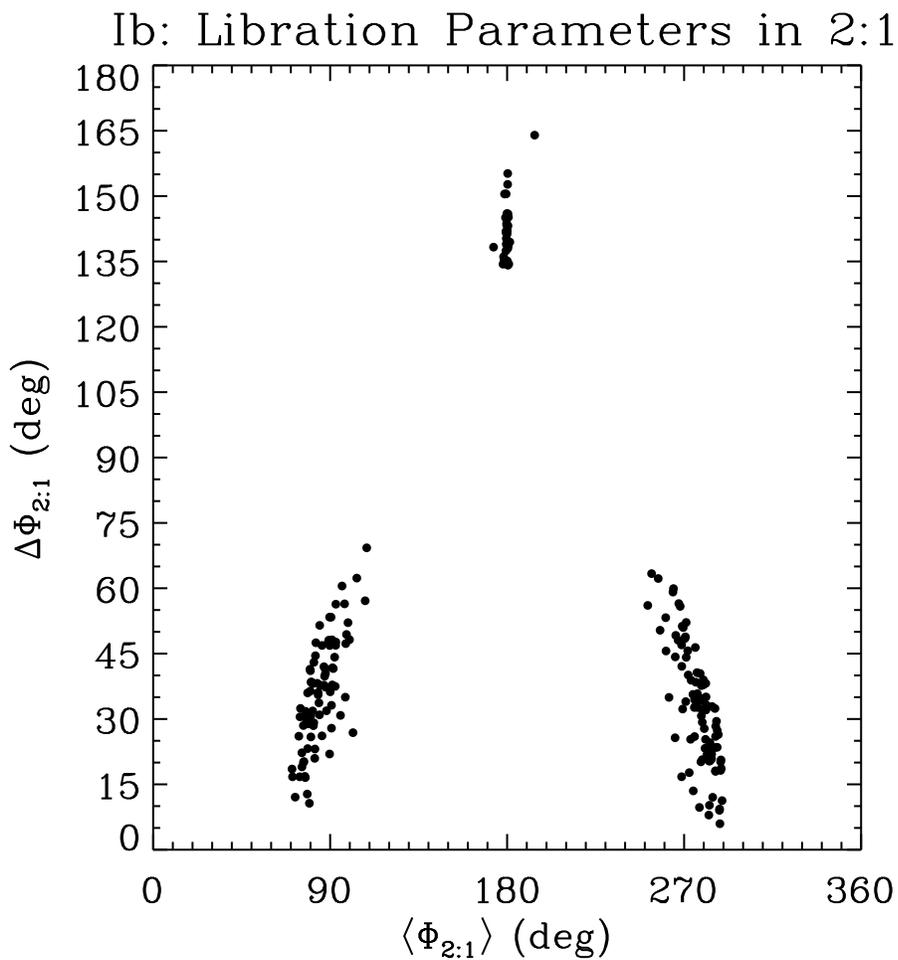}
\caption{Amplitude of libration vs.~libration center for Twotinos.
The majority of Twotinos
librate about $\langle \Phi_{2:1} \rangle \approx \pm \pi/2$ and have
libration amplitudes that are smaller than those of Twotinos
librating about $\langle \Phi_{2:1} \rangle = \pi$.
The lobe at $\langle \Phi_{2:1} \rangle \approx 3\pi/2$
contains 10\% more particles
than the lobe at $\langle \Phi_{2:1} \rangle \approx \pi/2$; this asymmetry
becomes more pronounced as shorter migration timescales are considered;
compare with Figure \ref{lib21IIb}.
\label{lib21}}
\end{figure}

For the Twotinos in simulation Ib, $\langle \Phi_{2:1} \rangle$
groups about 3 values: $\sim$$\pi/2$, $\pi$, and $\sim$$3\pi/2$.
The splitting of libration centers at large $e$ from
a single center at $\pi$ to two additional centers at $\sim$$\pi/2$
and $\sim$$3\pi/2$ was explored by Morbidelli, Thomas, \& Moons (1995),
Malhotra (1996), and references therein.
Figure \ref{lib21} plots $\Delta \Phi_{2:1}$ against
$\langle \Phi_{2:1} \rangle$. Objects that
reside more deeply in the resonance librate
about $\langle \Phi_{2:1} \rangle \approx 3\pi/2$ and $\pi/2$.
The former center is more heavily populated than the latter
at the level of 93 to 85 objects.
Though this difference
is formally statistically insignificant,
we nonetheless believe that it
reflects a heretofore unnoticed and physically
significant signature of the migration model.
Further evidence supporting
our contention is provided in \S\ref{vary}, where we demonstrate
that faster migration speeds dramatically enhance the asymmetry
to levels of statistical significance.
Unlike for the case of the 3:2 resonance, long-term, systematic
studies of the stability of objects within the 2:1 resonance have not
been published.
It might be thought that the requirements for stability within the 2:1
resonance are less stringent than for the 3:2 resonance, because
the $\nu_8$ and $\nu_{18}$ secular resonances do not
overlap the 2:1 (Morbidelli et al.~1995).
Simulations by Malhotra
(2002, personal communication) indicate that
the retainment efficiency is roughly $g_{2:1}= 50\%$,
and we will adopt this value in this paper.

\placefigure{fig7}

\begin{figure}
\epsscale{0.8}
\plotone{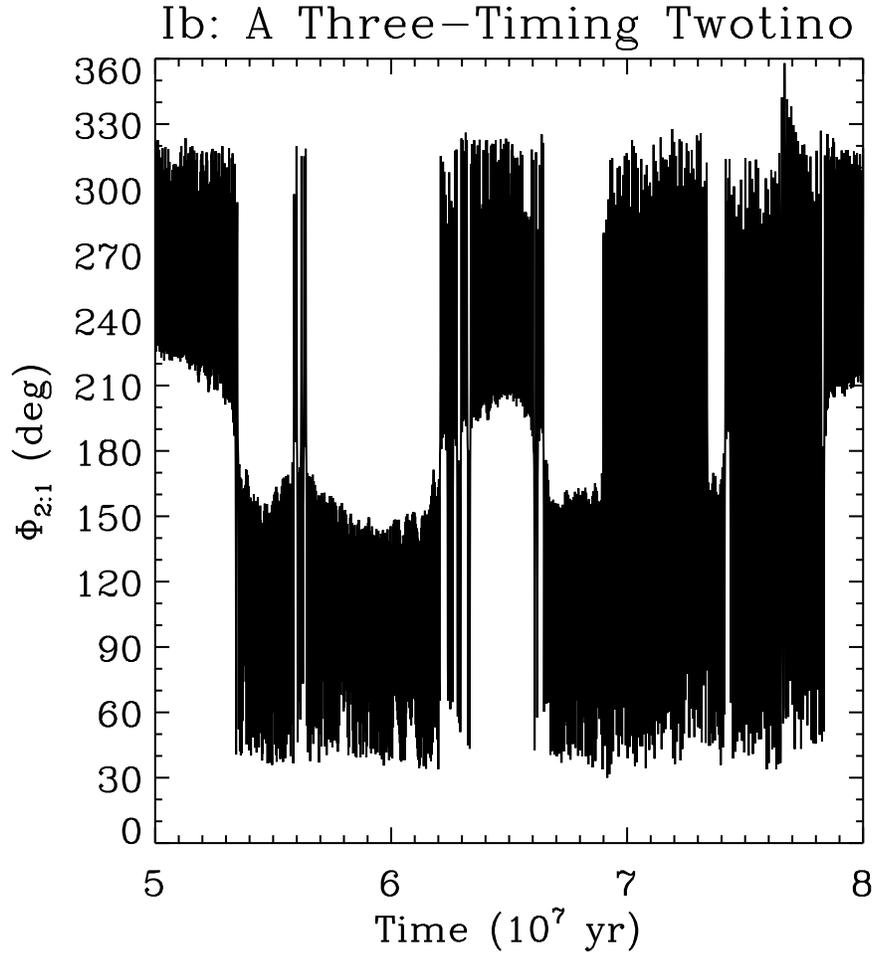}
\caption{A Twotino that alternates between librating about $\langle \Phi_{2:1}
\rangle \approx \pi/2$, $\pi$, and $3\pi/2$.
\label{threetimer}}
\end{figure}

\begin{figure}
\epsscale{0.8}
\plotone{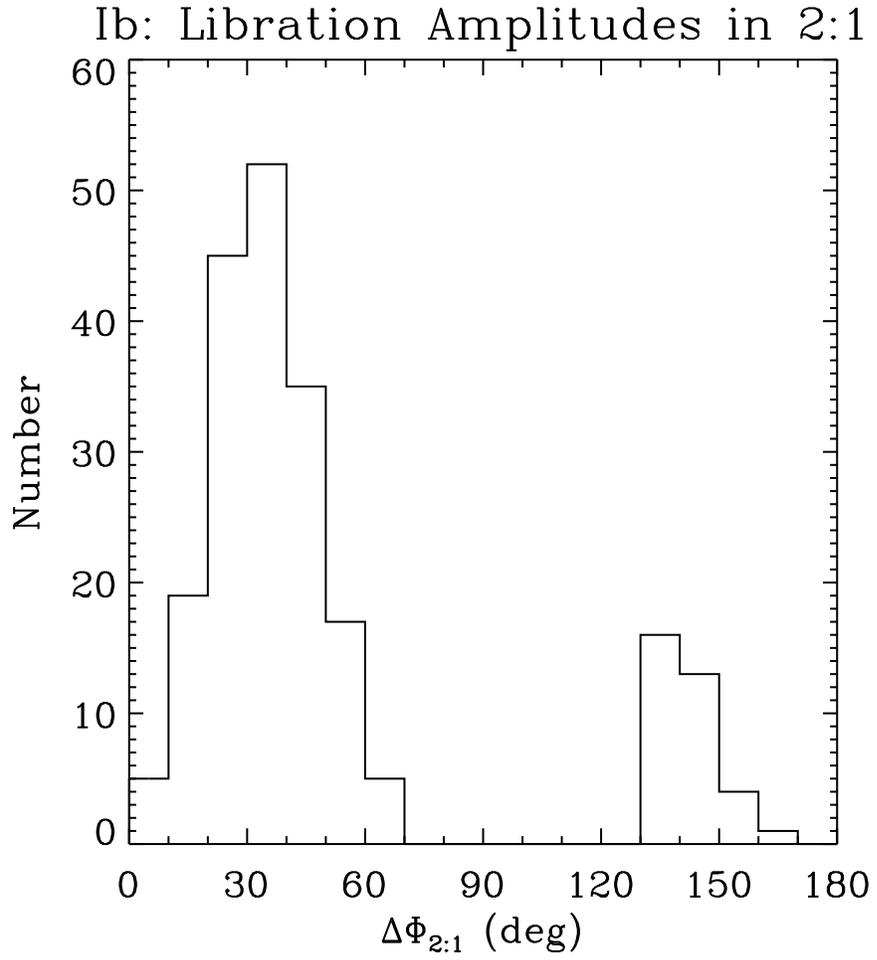}
\caption{Distribution of libration amplitudes of Twotinos in
simulation Ib.
\label{histo21}}
\end{figure}

Many of the objects having $\langle \Phi_{2:1} \rangle \equiv
[\max(\Phi_{2:1}) + \min(\Phi_{2:1})]/2 \approx \pi$ also
librate from time to time about $\sim$$\pi/2$ and $\sim$$3\pi/2$.
An example of such a ``three-timing'' Twotino is displayed
in Figure \ref{threetimer}. A histogram of libration amplitudes
of Twotinos is given in Figure \ref{histo21}.

\placefigure{fig8}

\placefigure{fig9}

\begin{figure}
\epsscale{0.8}
\plotone{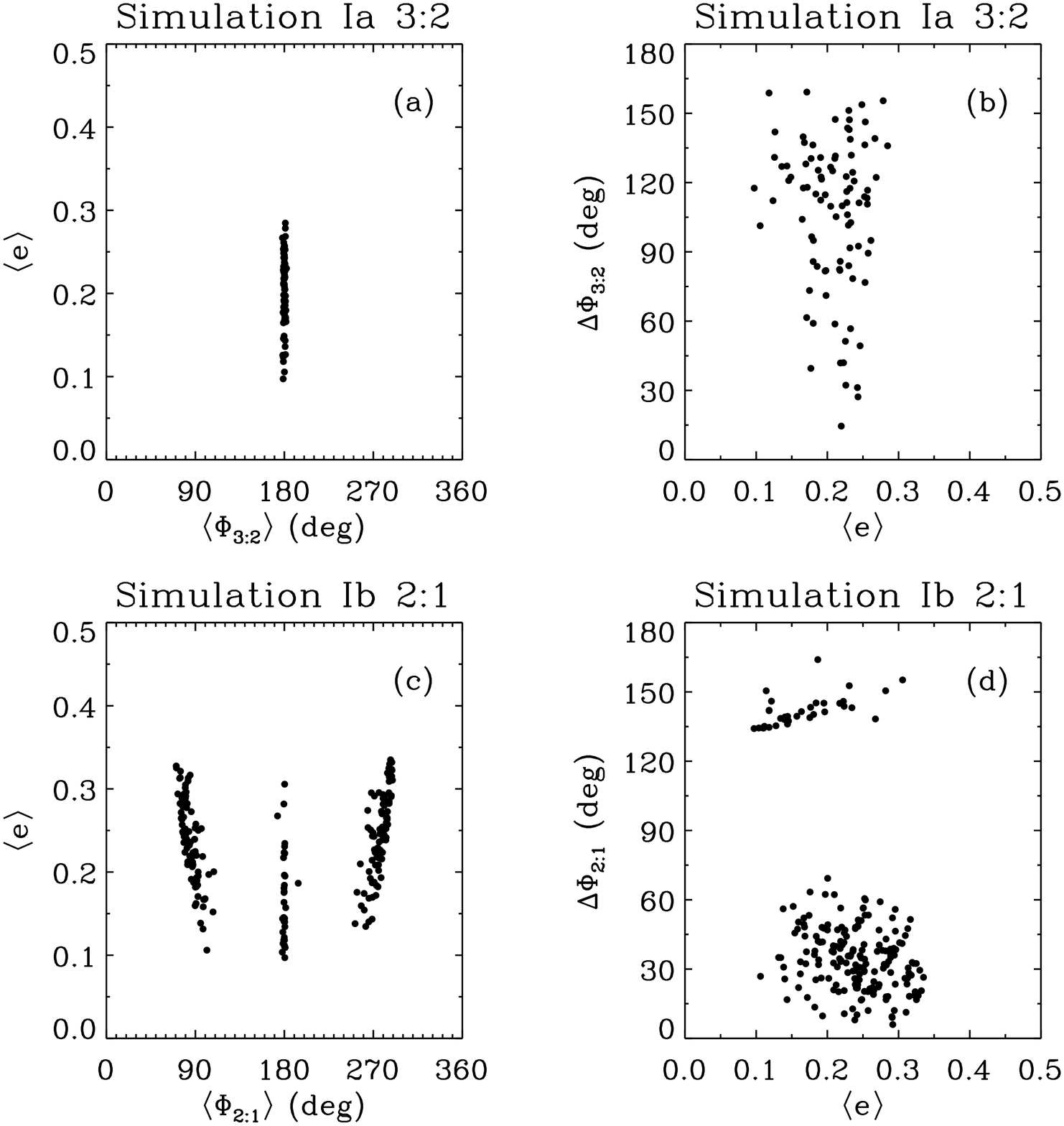}
\caption{(a) Average $e$ vs.~average $\Phi$ for Plutinos.
(b) Amplitude of libration vs.~average $e$ for Plutinos.
(c,d) Same as (a,b) but for Twotinos.
As seen in (c), Twotinos with higher eccentricities librate
about centers closer to $\langle \Phi_{2:1} \rangle = 0\degr$ than
those with lower eccentricities.
\label{ephi}}
\end{figure}

\begin{figure}
\epsscale{0.8}
\plotone{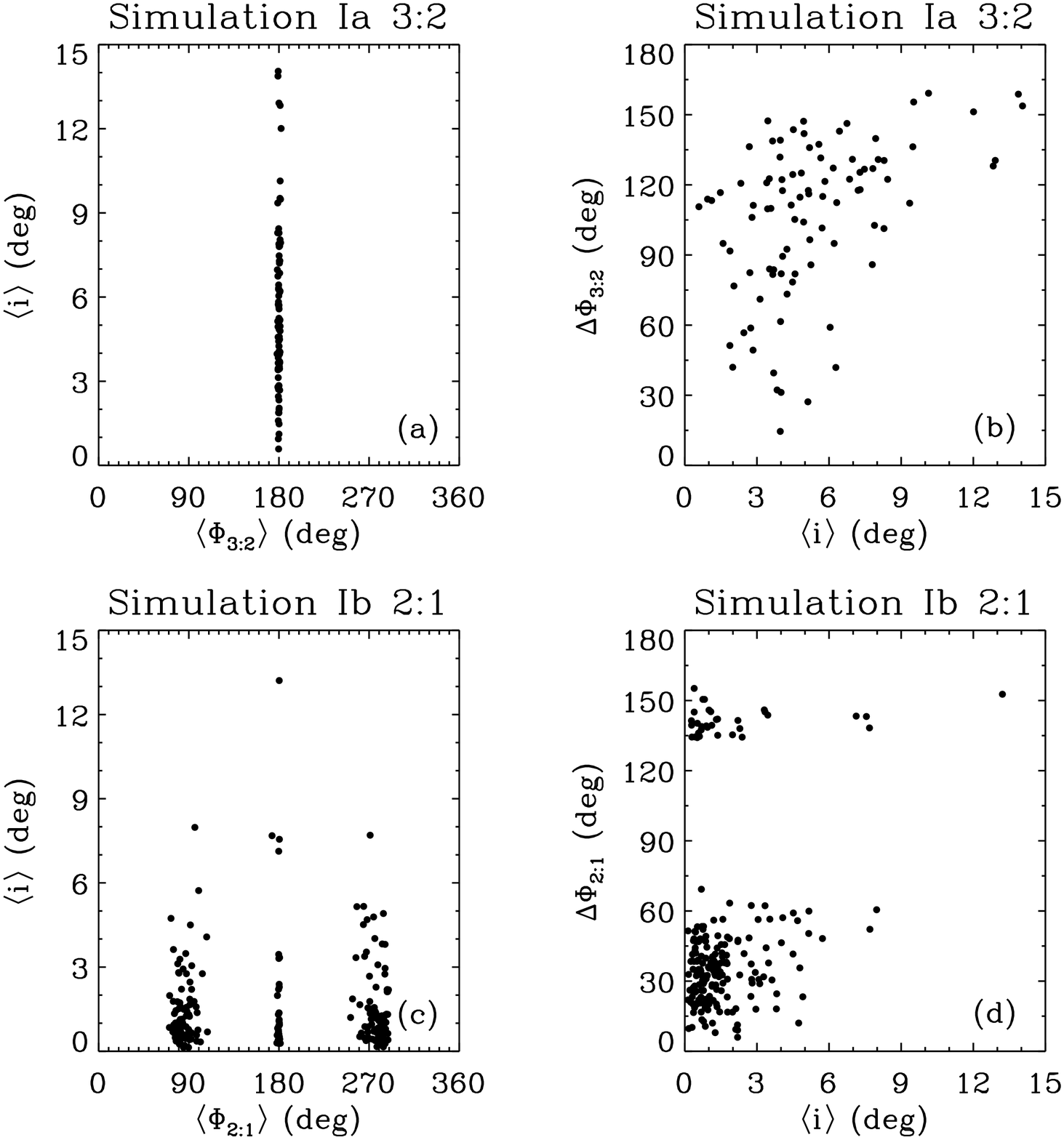}
\caption{(a) Average $i$ vs.~average $\Phi$ for Plutinos.
(b) Amplitude of libration vs.~average $i$ for Plutinos.
(c,d) Same as (a,b) but for Twotinos.
As seen in (b), highly inclined
Plutinos tend to be weakly bound to the first-order, eccentricity resonance.
\label{iphi}}
\end{figure}

For completeness, $\langle e \rangle$, $\langle i \rangle$,
$\langle \Phi_{j+1:j} \rangle$, and $\Delta \Phi_{j+1:j}$
are plotted against one another in Figures \ref{ephi}
and \ref{iphi}. One correlation that emerges is that
between $\langle e \rangle$ and $\langle \Phi_{2:1} \rangle$
for $\langle \Phi_{2:1} \rangle \approx \pi/2$, $3\pi/2$.
Increasing $\langle e \rangle$ increases the separation
of the two principal libration centers away from
$\langle \Phi_{2:1} \rangle = \pi$,
an effect seen previously by Malhotra (1996). Another correlation
appears between $\Delta \Phi_{3:2}$ and $\langle i \rangle$.
Large amplitude librators tend to have large $\langle i \rangle$.

\placefigure{fig10}

\placefigure{fig11}

\subsection{Capture into Secular Resonances}
\label{secresres}

\placefigure{fig12}
\begin{figure}[ph]
\epsscale{0.8}
\plotone{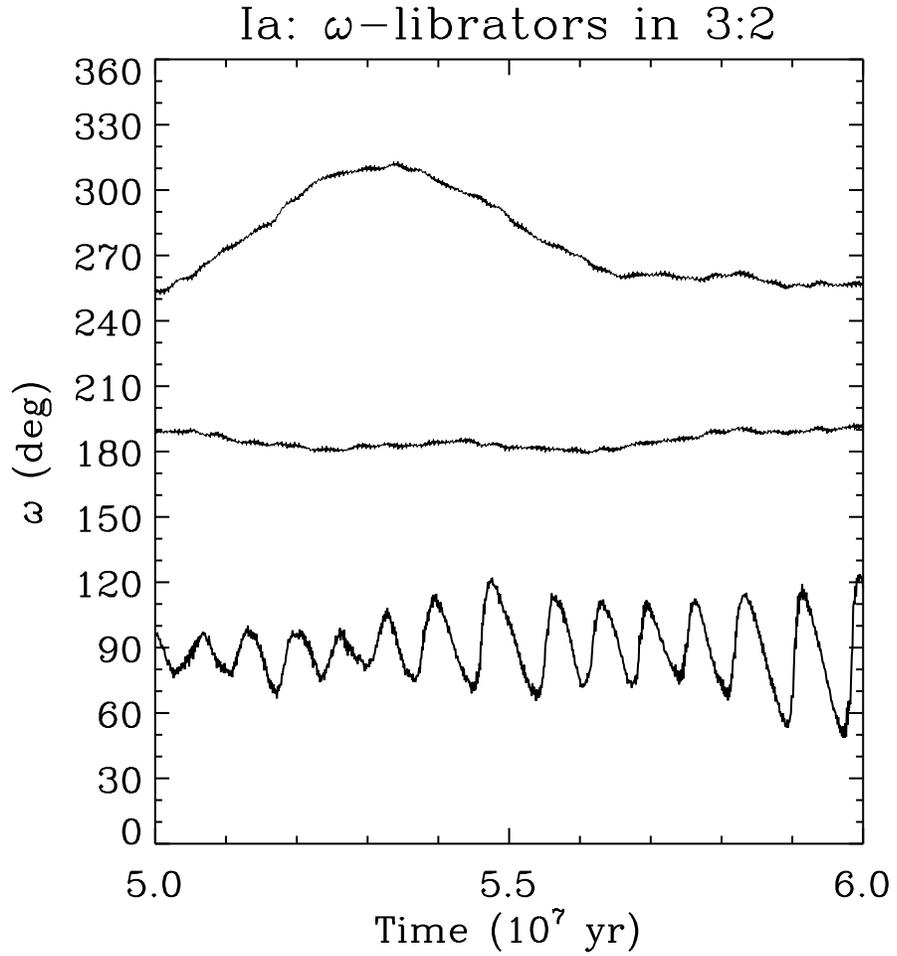}
\caption{Time evolution of $\omega$ for three Plutinos that also inhabit
Kozai-type resonances. Since the center of libration, $\langle \omega \rangle$,
does not take a unique value, and since fewer than 30\% of Plutinos
are found in the simulation to be $\omega$-librators, we conclude that
Kozai-type resonances do not introduce strong latitudinal biases
for finding Plutinos.
\label{w32}}
\end{figure}

\placefigure{fig13}
\begin{figure}[ph]
\epsscale{0.8}
\plotone{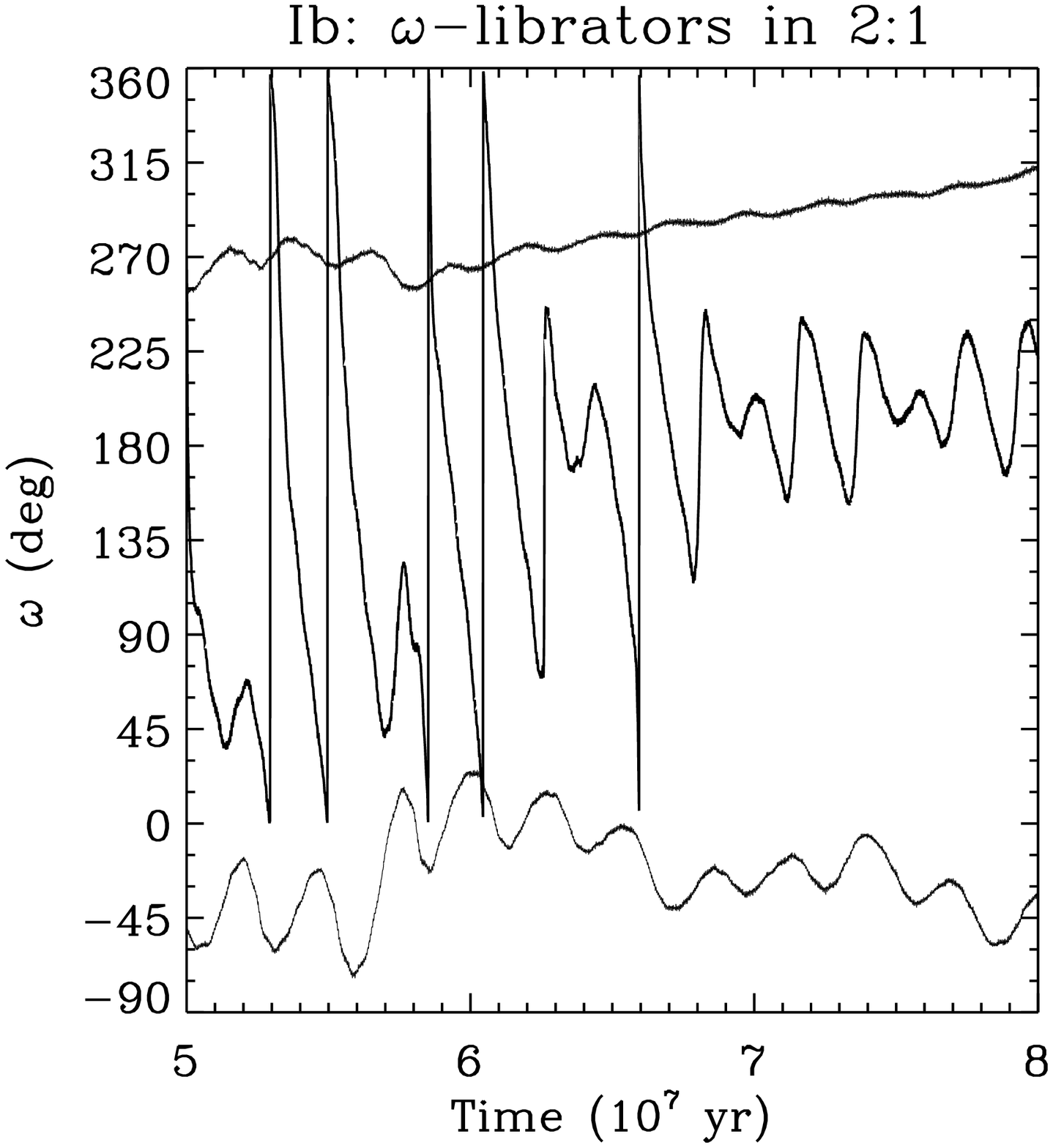}
\caption{Time evolution of $\omega$ for three Twotinos in Kozai-type
resonances.
Fewer than 7\% of the Twotinos in simulation Ib are $\omega$-librators.
\label{w21}}
\end{figure}

Of the 92 captured Plutinos in simulation Ia, 9--17 evince libration
of $\omega$ over the last $1\times 10^7\yr$ of the simulation.
Our uncertainty arises because for 8 Plutinos,
the duration of the simulation is too short to see either
a complete cycle of libration or of circulation. The libration
centers, $\langle \omega \rangle$, are distributed over
$\sim$90$\degr$, $\sim$$180\degr$, and
$\sim$$270\degr$ for the 9 confirmed $\omega$-librators.
A sampling of the time evolution
of $\omega$ for three $\omega$-librators in the 3:2 resonance
is provided in Figure \ref{w32}.

Only a subset of the 92 captured Plutinos are likely to remain bound
to the 3:2 resonance over $4\times 10^9\yr$. Of the 92, we
estimate that 42 are potential long-term residents: they either
have $\Delta \Phi_{3:2} < 110\degr$ or their $\omega$'s librate
with small amplitude about $\pm 90\degr$
[compare with Levison \& Stern (1995)]. Of these 42, 8--12 are
$\omega$-librators, or 20--30\%.
Since the $\omega$-librators comprise a minority among surviving
3:2 resonant objects, and since $\langle \omega \rangle$
can equal $180\degr$ in addition to $\pm 90\degr$, we conclude
that secular libration of $\omega$ resulting from the migration model
probably does not introduce strong
latitudinal selection effects for the discovery of Plutinos.
Of course, a zeroth-order assessment of
latitudinal selection effects for resonant objects probably requires
that we move beyond, or in the extreme case abandon, the model
for resonant migration, since it apparently cannot reproduce very well
the observed inclination distributions; see \S\ref{capeff}
and Brown (2001).

Similar conclusions are obtained for Twotinos.
Of the 212 objects captured into the 2:1 resonance,
only 12--15 evince libration of $\omega$. Figure
\ref{w21} samples three of them.

Our results address a question posed by Nesvorny, Roig, \& Ferraz-Mello (2000;
see their footnote 3). Could the observed paucity of Plutinos in the
Kozai resonance today be a primordial relic of resonant capture
and migration? Of the 42 captured Plutinos in simulation Ia
that might survive in the 3:2 resonance over the age of the solar
system, only 3--5 occupy a Kozai resonance for which
$\langle \omega \rangle = 90\degr$. This fraction
of $\sim$4/42 compares well with the observed fraction of $\sim$3/33 reported
by Nesvorny et al.~(2000). Thus, our answer to their question is
yes---scattering effects by Pluto need not be invoked to explain
today's observed low fraction of Plutinos in the Kozai resonance.
We leave unaddressed two issues: (1) the effects of collisions
amongst KBOs in the primordial belt in populating the Kozai resonance,
and (2) whether the fact that the most massive Plutino yet discovered
also inhabits the Kozai resonance is coincidental.

\subsection{Spatial Distribution of Resonant Objects}
\label{snap}

We synthesize instantaneous snapshots of the spatial distribution
of resonant objects from our simulation data as follows. Essentially,
the positions of $R$ resonant particles sampled at $T$ different
times are taken to represent the positions of $R \times T$ particles
sampled at 1 time. Take the 2:1 resonance as an example. The inertial
Cartesian coordinates of Neptune $[(x_N, y_N, z_N)]$ and of the
$R = 212$ Twotinos $[(x_i, y_i, z_i), i=\{1,\ldots,212\}]$ at
a time, $t$, in the simulation are rotated about the $\hat{z}$-axis
by an angle

\begin{equation}
\Theta(t) = \arccos{\frac{x_N (t) X_N + y_N (t) Y_N}{\sqrt{[x_N(t)^2 +
y_N(t)^2] (X_N^2 + Y_N^2)}}} \, .
\end{equation}

\noindent Here $X_N = 30.1 \cos{(302\degr)} \AU$ and
$Y_N = 30.1 \sin{(302\degr)} \AU$
are the approximate current coordinates of Neptune.
This procedure shifts the positions of all bodies into the
Neptune-centric frame. We repeat
this operation for $T$ different instants of the simulation to generate
$T$ different snapshots. These snapshots are then overlaid on one another
to yield a single image, a representation of the present-day
spatial distribution of $R\times T$ resonant particles.

\begin{figure}
\epsscale{0.8}
\plotone{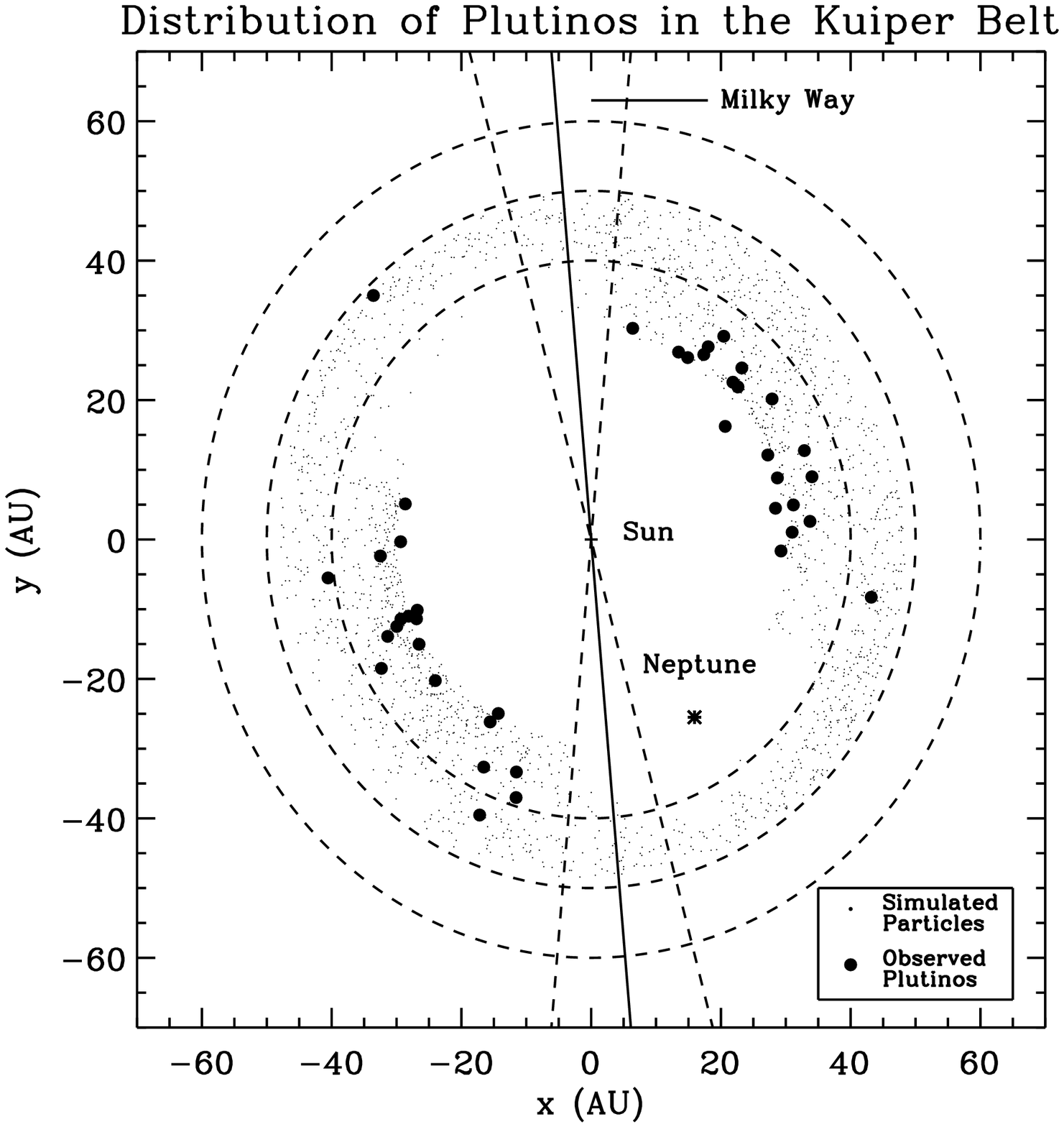}
\caption{Synthesized snapshot, viewed from the invariable pole,
of the spatial distribution of low libration
amplitude Plutinos in simulation Ia, for which $\tau=10^7\yr$.
Large black dots mark the positions of observed Plutino candidates
catalogued by the Minor Planet Center as of July 4, 2002.
The main features of the snapshot, including the ``sweet
spot'' concentrations of Plutinos displaced
$\pm 90\degr$ from Neptune's longitude,
and the relative dearth of Plutinos at longitudes $0\degr$ and 180$\degr$
from Neptune's, can be reproduced by a simple
toy model; compare with Figure \ref{toy32}b.
Dashed circles indicate heliocentric distances of 40, 50, and 60 AU.
Radial lines delineate where the Galactic plane, $\pm 10\degr$ Galactic
latitude, intersects the invariable plane; Kuiper Belt surveys avoid
the Galactic plane because fields there tend to be too crowded with stars.
\label{snap32}}
\end{figure}

As noted in \S\ref{libstat}, only a subset of captured Plutinos
will be retained by the 3:2 resonance over $4\times 10^9\yr$.
We employ the subset of $R = g_{3:2}\times f_{3:2}\times 400 = 42$
objects to construct
the Plutino snapshot plot (Figure \ref{snap32})
and those figures quantifying
the longitudinal variations of Plutino density
(Figures \ref{objperdeg}a and \ref{dobjperdeg}).

Our method is not strictly justifiable if Neptune does not
execute a perfectly circular orbit that remains in the invariable
plane. In practice, Neptune's eccentricity and
inclination are so small that we do not consider this a
serious violation. A more weighty concern is whether the
true distributions of orbital elements---eccentricities,
inclinations, and libration amplitudes---of $R \times T$
particles are well represented by only $R$ particles.
Since $R$ is not small for either simulation, and
since the distribution functions displayed in
Figures \ref{histo32allin1}, \ref{lib21}, \ref{histo21},
\ref{ephi}, and \ref{iphi} do not betray poor coverage
of phase space, we proceed with confidence.

\begin{figure}
\epsscale{0.8}
\plotone{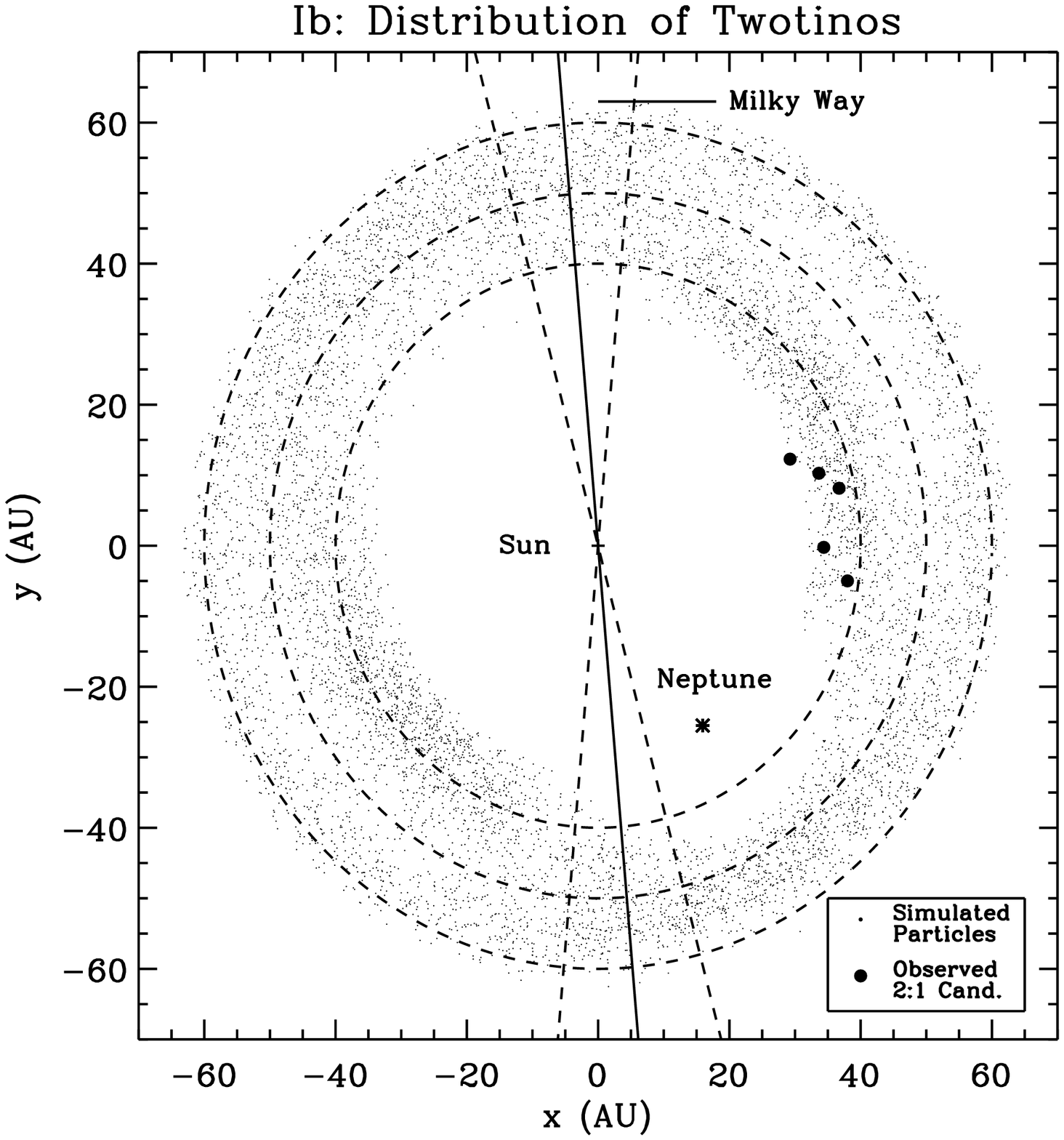}
\caption{Synthesized snapshot, viewed from the invariable pole,
of the spatial distribution of Twotinos
subsequent to resonant capture in simulation Ib, for which $\tau=10^7\yr$.
Large black dots mark the positions of observed Twotino candidates
catalogued by the Minor Planet Center as of July 4, 2002.
The main features of the snapshot, including the ``sweet
spot'' concentrations of Twotinos displaced
$\pm 75\degr$ from Neptune's longitude,
and the relative dearth of Twotinos at longitudes near
those of Neptune, can be reproduced by a simple
toy model; compare with Figure \ref{toy21}d.
Dashed circles indicate heliocentric distances of 40, 50, and 60 AU,
and radial lines delineate where the Galactic plane, $\pm 10\degr$
Galactic latitude, intersects the invariable plane.
Strangely, though the model predicts roughly equal numbers
of Twotinos ahead and behind of Neptune's longitude,
Twotino candidates have only been detected in the former lobe.
Contrast this snapshot with the one in Figure \ref{snap21longIIb},
where we consider a migration phase that is $10 \times$ shorter.
\label{snap21}}
\end{figure}

\begin{figure}
\epsscale{0.8}
\plotone{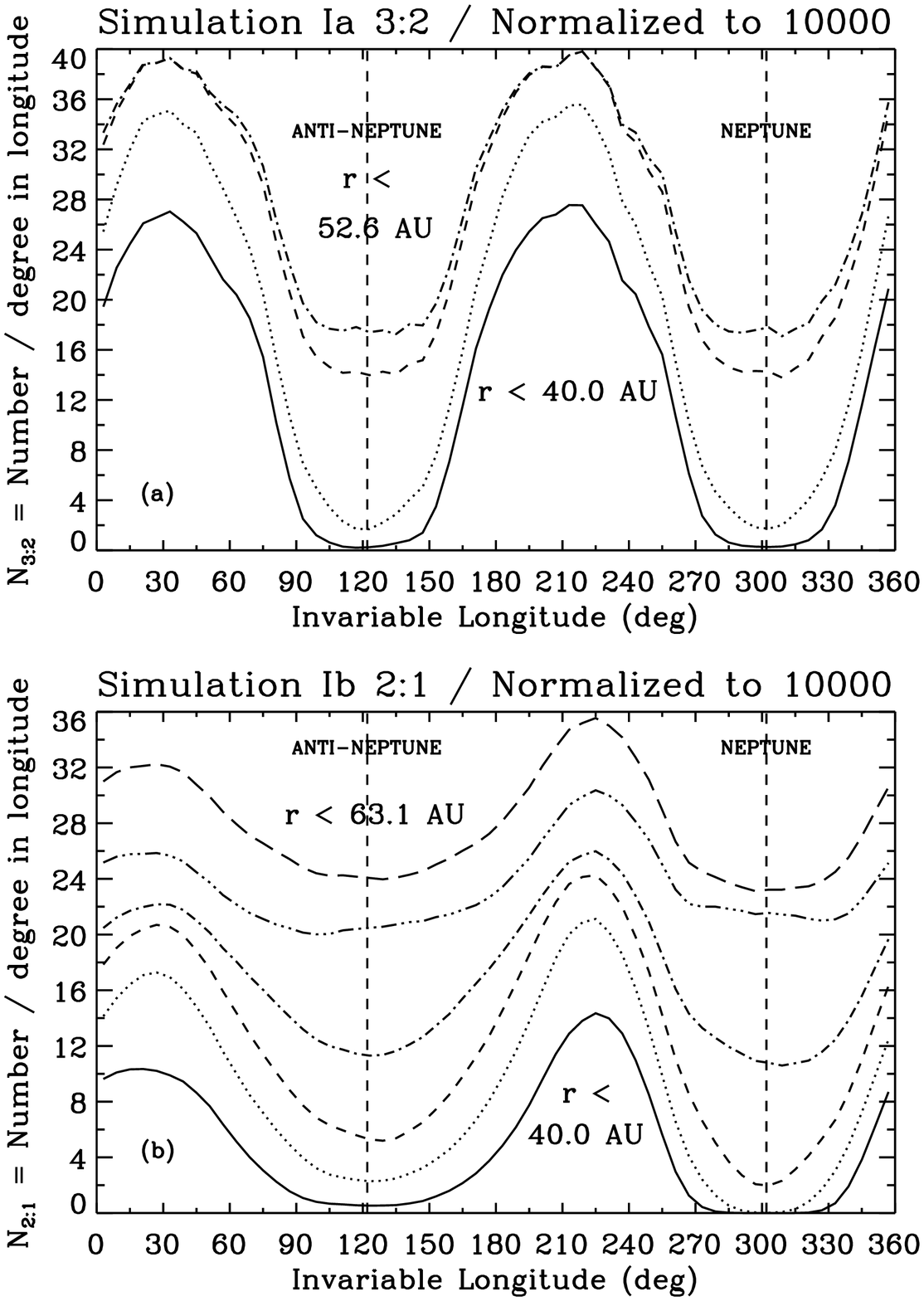}
\caption{Longitudinal distribution of
(a) Plutinos selected for likely long-term
stability, and (b) Twotinos. Data are synthesized from
simulation I, for which the migration timescale $\tau = 10^7\yr$.
The number of objects within each resonance, integrated over all longitudes,
is normalized to 10000. Each curve represents those
objects located at heliocentric distances, $r$, less than the
value shown. Adjacent curves differ by a multiplicative factor
of 1.0955 in limiting $r$. Longitudes for Neptune and for the anti-Neptune
direction are indicated. The contrast in object
density from longitude to longitude is
greatest at small limiting $r$. Plutinos cluster at
longitudes $\pm$90$\degr$ away from Neptune, while
Twotinos are most likely found $\pm$75$\degr$ away.
\label{objperdeg}}
\end{figure}

\begin{figure}
\epsscale{0.8}
\plotone{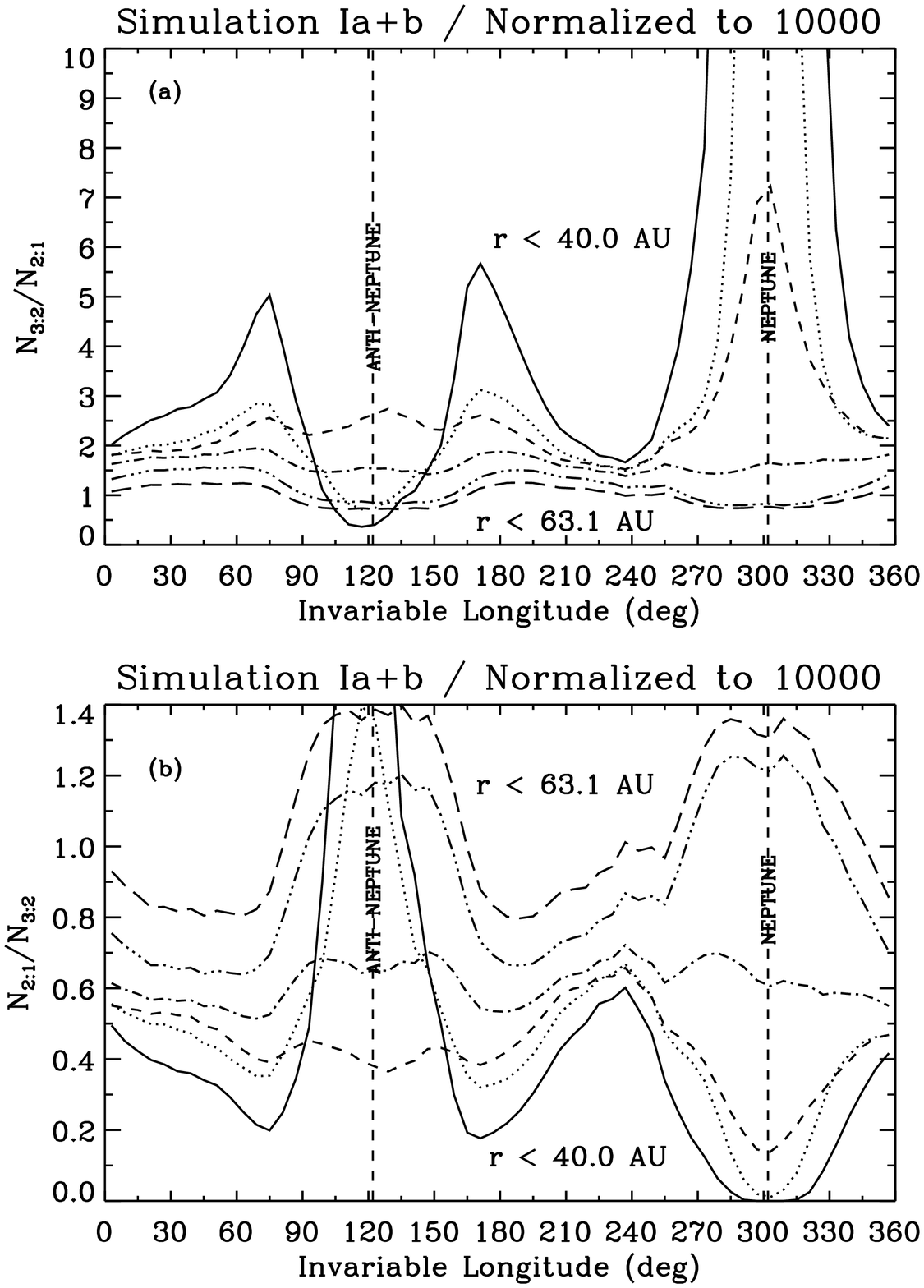}
\caption{Longitudinal variations of the bias in finding Twotinos
vs.~Plutinos in simulation I, for which $\tau=10^7\yr$. As in Figure
\ref{objperdeg},
the number of objects in each resonance is normalized to 10000.
Panel (a) plots the fraction of Plutinos to Twotinos found
inside a heliocentric distance, $r$, while panel (b) plots
the inverse ratio. Adjacent curves differ by a factor of 1.0955 in limiting
$r$. At longitudes between $210\degr$ and $240\degr$
(between $-90\degr$ and $-60\degr$ of Neptune's longitude),
yields of Twotinos and Plutinos are expected to be nearly equal.
A similar interval exists between $0\degr$ and $40\degr$ longitude
(+$60\degr$--$100\degr$ of Neptune).
\label{dobjperdeg}}
\end{figure}

Figures \ref{snap32} and \ref{snap21} showcase the present-day
snapshots of Plutinos and Twotinos, respectively. For the former
figure, $T = 1000$ time slices of $R=42$ particles
are sampled uniformly between
$t_1 = 5 \times 10^7 \yr$ and $t_2 = 6 \times 10^7 \yr$;
for the latter figure, $T = 190$, $R=212$, $t_1 = 5.000 \times 10^7 \yr$,
and $t_2 = 5.190 \times 10^7 \yr$. More time slices could be
sampled for the Twotinos, but we restrict $T$ to keep the plots legible.
These snapshots resemble closely the toy models presented
in Figures \ref{toy32}b and \ref{toy21}d; we conclude that
(weak) correlations between orbital elements such as $\Delta \Phi_{2:1}$
and $\langle e \rangle$---correlations that are missing
from the toy models---do not significantly influence the organization
of resonant populations.

\placefigure{fig14}

\placefigure{fig15}

Plutinos in the migration model cluster
$\pm 90\degr$ away from Neptune. Twotinos
cluster $\pm 75\degr$ away. Plutinos avoid longitudes
near Neptune and longitudes
that are $180\degr$ away from Neptune. Twotinos
largely avoid longitudes near Neptune. Longitudinal
variations in the density of resonant objects are
quantified in Figure \ref{objperdeg},
where we plot the number of objects per degree in longitude, $N_{j+1:j}$,
inside a heliocentric distance, $r$. In constructing
Figure \ref{objperdeg}, we employ all available time slices ($T=1000$
in simulation Ia and $T=3000$ in simulation Ib), and then normalize the curves
for $N_{j+1:j}$ so that the total population within each resonance,
integrated over all longitudes, equals 10000 objects.

\placefigure{fig16}

The sweet spots on the sky for finding resonant objects are sweetest---that
is, the contrast between maximum object density and minimum object
density is greatest---for small limiting
$r$ (e.g., $r \leq 40 \AU$). In the large $r$ limit, the maximum
contrast in object density is $\sim$200\% for Plutinos and $\sim$40\%
for Twotinos. Note that the two sweet spots for Twotinos differ
in strength; the spot displaced by $-75\degr$ from Neptune
contains more objects than the spot displaced by $+75\degr$, reflecting
a greater population of objects librating about
$\langle \Phi_{2:1} \rangle \approx 3\pi/2$ rather than $\sim$$\pi/2$.
See \S\ref{libstat} and \S\ref{vary} for more discussion of this asymmetry.

\placefigure{fig17}

In Figure \ref{dobjperdeg}, we divide $N_{3:2}$ by $N_{2:1}$
to compute the longitudinal bias in finding Plutinos over
Twotinos. Though the population
of each resonance is normalized to the same number, many more
Plutinos will be found than Twotinos at small $r$, simply
because the 3:2 resonance is located closer than the 2:1. Most interestingly,
however, there exists a special longitude interval between $210\degr$
and $240\degr$ (between $-90\degr$ and $-60\degr$ of Neptune's
longitude) where approximately equal numbers of Plutinos and Twotinos
are expected to be found. Within this longitude interval,
the yield of Twotinos to Plutinos ranges from 0.4 to 1 as
the limiting $r$ increases from 40 AU to $\infty$ (assuming that
the two resonances are equally well populated). The absolute
object density ($N_{j+1:j}$) for each resonance is also maximal
over this longitude range, making this interval the sweetest of spots.
A similar spot exists between $0\degr$ and $40\degr$ longitude
($+60\degr$--$100\degr$ of Neptune's longitude); here
$N_{2:1}/N_{3:2}$ varies from 0.3 to 0.9 as the limiting
$r$ increases from 40 AU to $\infty$.

\section{VARYING THE MIGRATION SPEED}
\label{vary}

Simulations IIa and IIb are identical to Ia and Ib, respectively,
except that we set $\tau = 10^6\yr$ and $t_f^{\rm{II}} = 5\times 10^6\yr$.
For simulations IIIa and IIIb,
$\tau=10^5\yr$ and $t_f^{\rm{III}} = 5 \times 10^5\yr$.

Table \ref{capefftab} summarizes the computed capture efficiencies, $f$,
and estimated retainment efficiencies, $g$, for each simulation.
The retainment efficiency equals the fraction of captured
objects that are expected to remain in the resonance
over $4\times 10^9\yr$.
For Twotinos, we assume that $g=0.5$ (see \S\ref{libstat}).
For Plutinos, $g$ equals the fraction of captured objects
that either have $\Delta \Phi < 110\degr$ or
that librate about $\langle \omega \rangle = \pm 90\degr$.

For $\tau=10^6\yr$, $f_{2:1} \approx 15$\%,
more than three times lower than the corresponding value for
$\tau = 10^7\yr$. By contrast, $f_{3:2} (\tau = 10^6\yr) \approx 78\%$,
more than three times as high as $f_{3:2} (\tau = 10^7\yr)$ because
fewer objects are lost to capture by the 2:1 or to close
encounters with Neptune. For $\tau = 10^5\yr$, $f_{2:1} \approx 0\%$,
while $f_{3:2} \approx 30\%$. Our results are consistent
with those of Ida et al.~(2000) and Friedland (2001).

\placefigure{fig18}
\begin{figure}[p]
\epsscale{0.8}
\plotone{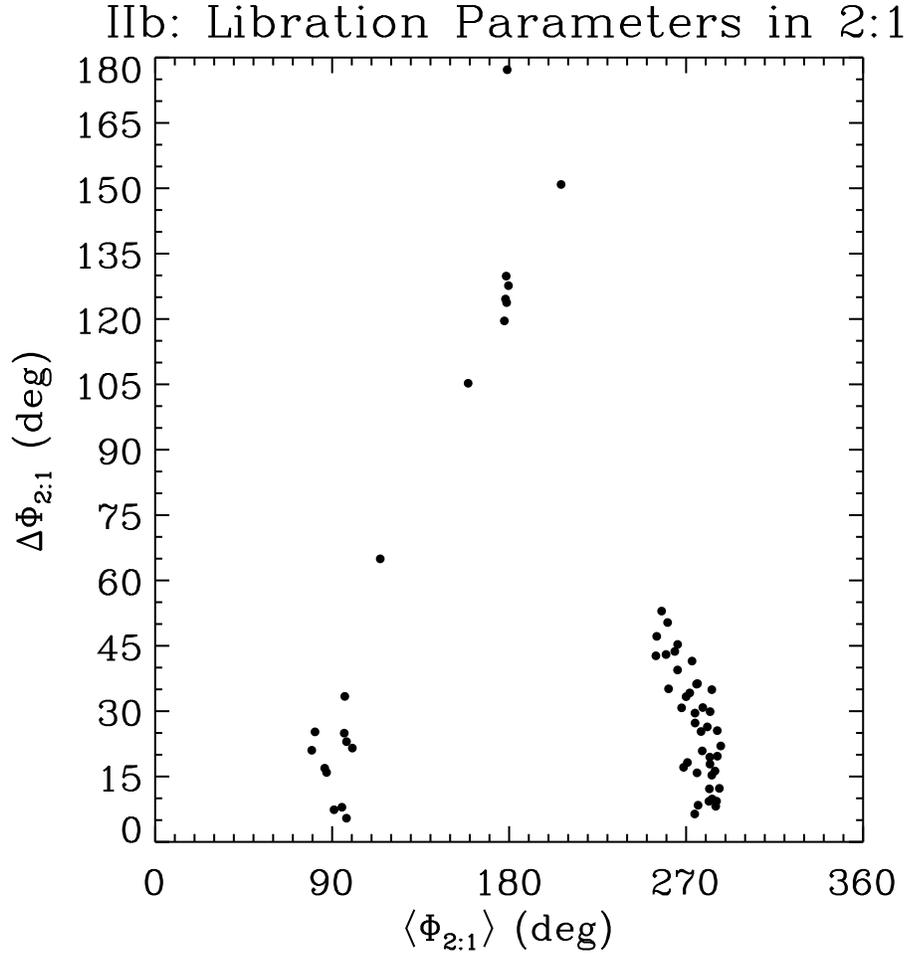}
\caption{Amplitude of libration vs.~libration center
for Twotinos in simulation
IIb, for which $\tau = 10^6\yr$. The lobe at
$\langle \Phi_{2:1} \rangle = 270\degr$ boasts greater membership
than the lobe at $\langle \Phi_{2:1} \rangle = 90\degr$ by
a factor of 3.3-to-1. Contrast this result with Figure \ref{lib21},
for which $\tau=10^7\yr$.
\label{lib21IIb}}
\end{figure}

\placefigure{fig19}
\begin{figure}[ph]
\epsscale{0.8}
\plotone{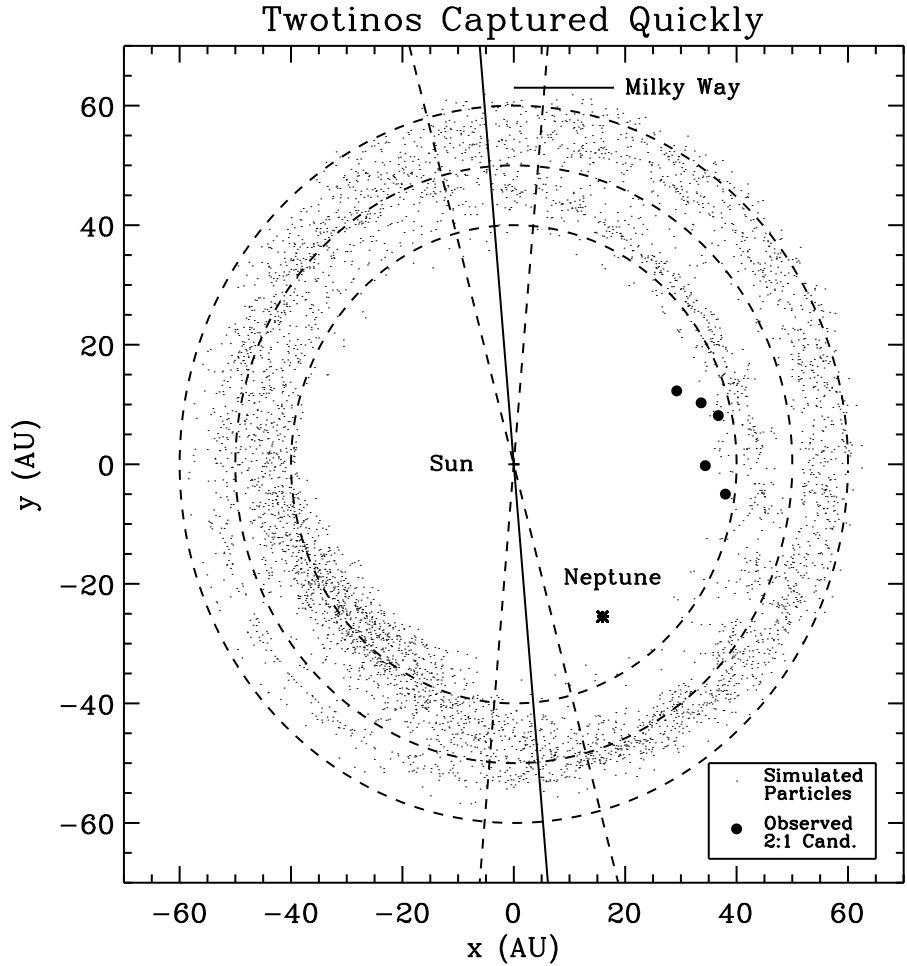}
\caption{Snapshot of the spatial distribution of Twotinos in simulation IIb,
viewed from the invariable pole.  Large black dots represent observed Twotino
candidates catalogued by the Minor Planet Center as of July 4, 2002.  The
shorter migration timescale of $\tau = 10^6\yr$
leads to a pronounced asymmetry of objects in
space. Dashed circles delimit heliocentric radii of 40, 50, and 60 AU,
and radial lines indicate the intersection of the
Galactic plane with the invariable plane.
\label{snap21longIIb}}
\end{figure}

\placefigure{fig20}
\begin{figure}[ph]
\epsscale{0.8}
\plotone{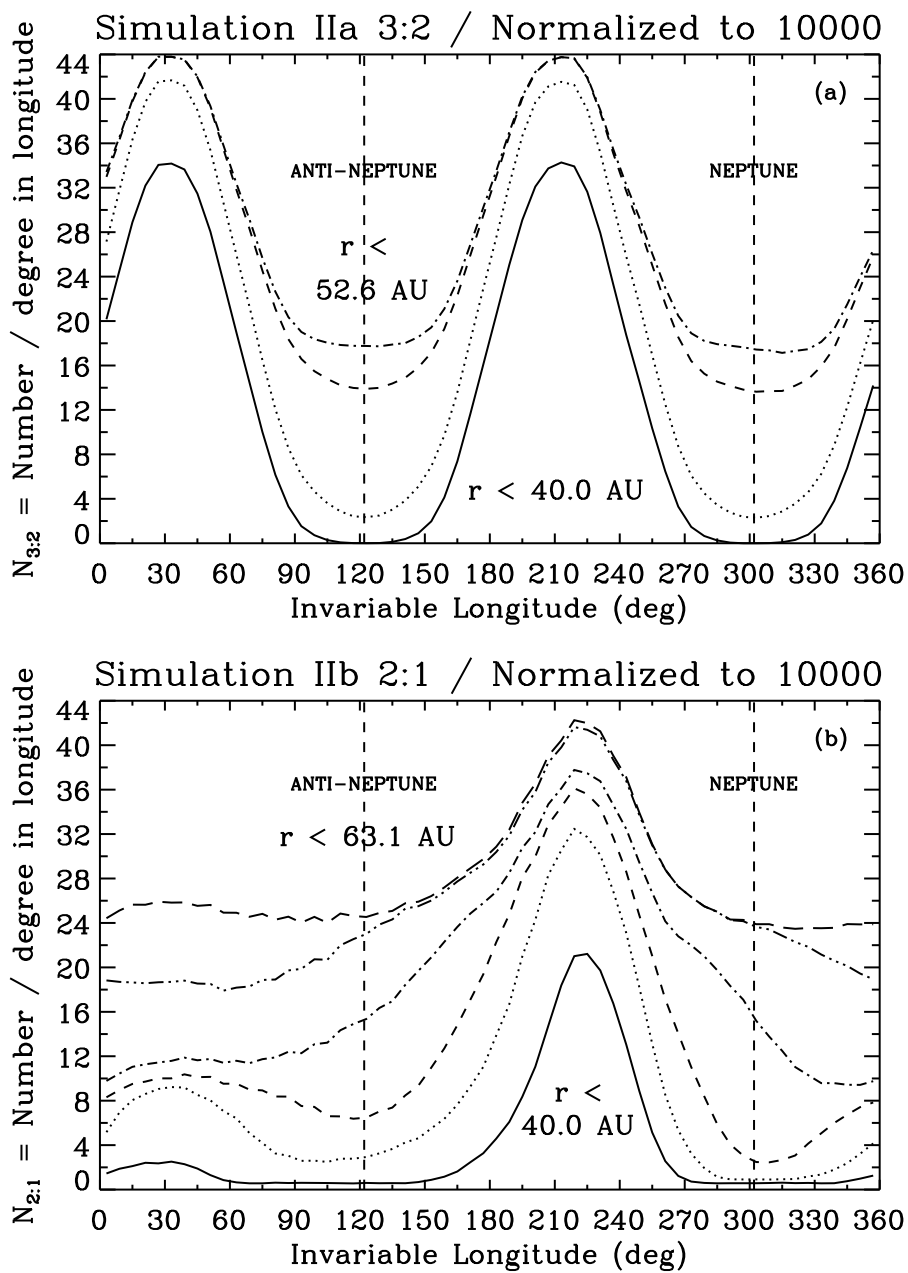}
\caption{Longitudinal distribution of (a) Plutinos and (b) Twotinos
in simulation II for which $\tau=10^6\yr$. The curves are
normalized so that each
resonance contains 10000 objects; adjacent curves differ
by a factor of 1.0955 in limiting $r$. For Twotinos, the sweet spot located
$-75\degr$ away from Neptune's longitude is substantially sweeter than
the one at $+75\degr$. Compare with Figure \ref{objperdeg}
which portrays the results of simulation I
for which $\tau=10^7\yr$.
\label{objperdegII}}
\end{figure}

\placefigure{fig21}
\begin{figure}[ph]
\epsscale{0.8}
\plotone{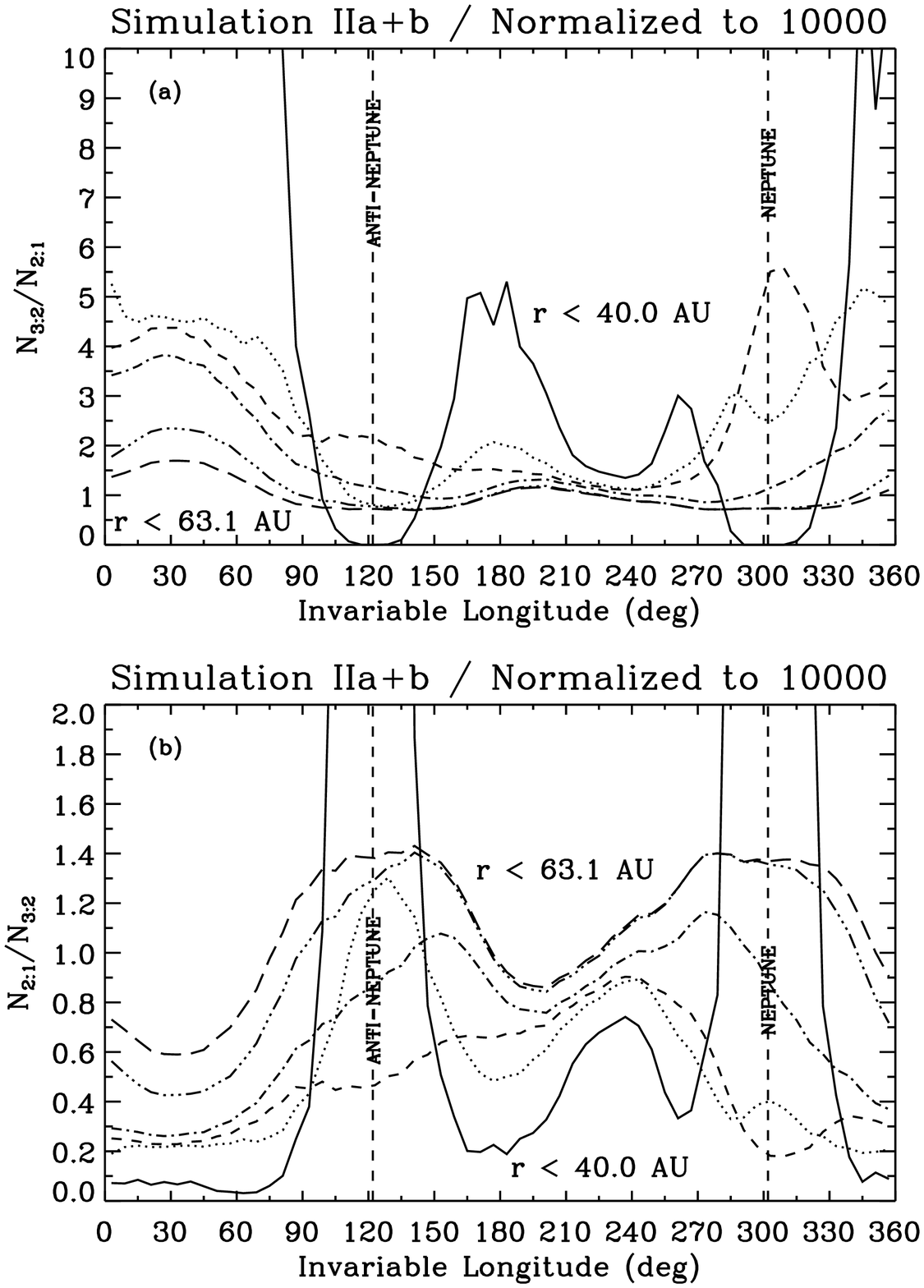}
\caption{Longitudinal variations of the bias in finding Twotinos
vs.~Plutinos in simulation II, for which $\tau=10^6\yr$.
As in Figure \ref{objperdegII}, the number of objects in
each resonance is normalized to 10000. By
contrast with simulation I for which $\tau=10^7\yr$, only
the interval between $210\degr$ and $240\degr$ longitude ($-90\degr$
to $-60\degr$ of Neptune's longitude) exhibits minimal bias
in favor of detecting Plutinos over Twotinos; there, the yield
of 3:2 resonant objects over 2:1 resonant objects ranges from 1 to 2.5
as the limiting heliocentric distance, $r$, out to which objects
are detected increases from 40 AU to $\infty$.
\label{dobjperdegII}}
\end{figure}

The Plutinos' distribution of libration amplitudes
shifts substantially towards smaller values as $\tau$ is
reduced. Figure \ref{histo32allin1} plots histograms of $\Delta \Phi_{3:2}$
for all three simulations, Ia, IIa, and IIIa. Table \ref{capefftab}
records how the retainment efficiency, $g$,
more than doubles for Plutinos as $\tau$ is
reduced from $10^7\yr$ to $10^6\yr$, a consequence
of the smaller libration amplitudes
that characterize faster migration rates.

Remarkably, objects fill the 2:1 resonance asymmetrically: captured
Twotinos prefer to librate about $\langle \Phi_{2:1} \rangle \approx 270\degr$
rather than $\langle \Phi_{2:1} \rangle \approx 90\degr$.
Figure \ref{lib21IIb} plots $\Delta \Phi_{2:1}$ against
$\langle \Phi_{2:1} \rangle$ for simulation IIb. The
preferential filling of one resonance lobe over another
dramatically affects the spatial distribution of Twotinos,
as illustrated by Figure \ref{snap21longIIb}
and as quantified in Figure \ref{objperdegII}.
The difference in populations is at the level of 330\% and is statistically
significant.\footnote{We have
verified that for $\tau = 3 \times 10^6\yr$, the sign of
the asymmetry remains the same and its magnitude is statistically
significant and intermediate between that of simulations Ib and
IIb---among the 141/400 objects captured into the 2:1 resonance,
$3$ $\times$ as many Twotinos librate about
$\langle \Phi_{2:1}\rangle \approx 270\degr$ than about
$\langle \Phi_{2:1} \rangle \approx 90\degr$.}
Establishing the relative populations of Twotinos ahead and behind of
Neptune would offer a powerful
constraint on the migration history of that planet.

Figure \ref{dobjperdegII} is appropriate for $\tau=10^6\yr$
and is analogous to Figure \ref{dobjperdeg}.

\section{DISCUSSION}
\label{conc}

Under the hypothesis of resonant capture, the number of Plutinos
having diameters greater than $s$, divided by the number of similarly
sized Twotinos, is given today by

\begin{equation}
F(\gtrsim s, \tau) \equiv \frac{f_{3:2}(\tau) \, g_{3:2}(\tau) \,
\eta_0(\gtrsim s, a\approx 35\AU)}{f_{2:1}(\tau) \, g_{2:1}(\tau) \,
\eta_0(\gtrsim s, a\approx 42\AU)} \, .
\label{intrinf}
\end{equation}

\noindent Here $\eta_0(\gtrsim s, a)$ equals the number of objects having
diameters greater than $s$ that occupied an annulus of heliocentric
radius $a$ and radial width $\approx 8\AU$ just prior to the era
of Neptune's migration. From our findings in this paper,
$(f_{3:2}g_{3:2})/(f_{2:1}g_{2:1})$ equals 0.40, 9.6, and $\infty$ for
$\tau(\yr) = 10^7, 10^6$, and $10^5$, respectively (see Table \ref{capefftab}).

We interpret the observation that the 2:1 resonance today contains
at least one (candidate) object having a large eccentricity
to imply that the migration timescale, $\tau$, cannot be as
low as $10^5\yr$. More refined estimates of $\tau$ can be made
by measuring the relative number of Twotinos whose
resonant arguments $\langle \Phi_{2:1} \rangle$ librate
about $\sim$$3\pi/2$ rather than about $\sim$$\pi/2$; i.e., the
relative number of Twotinos observed to reach perihelion
at longitudes behind of as opposed to in front of Neptune's
longitude.

Estimates for the intrinsic $F$ cannot be established without first
de-biassing $F_{obs}$. As Figures \ref{dobjperdeg} and equation
(\ref{intrinf}) attest, $F_{obs}$ depends not only
on $\tau$ and the ratio of primeval populations,
$\eta_0 (a \approx 35\AU)/\eta_0 (a\approx 42\AU)$,
but also on (1) the longitude of observation,
(2) the limiting magnitude of the observation,
(3) the relative size distributions of Twotinos and Plutinos,
(4) the relative albedo distributions,
(5) the relative inclination distributions,
and, because of consideration (5), (6) the latitude
of observation. For purposes of discussion,
let us assume that Plutinos and Twotinos follow
the same size, albedo, and inclination distributions,
so that we can focus on considerations (1) and (2) exclusively.

Roughly speaking, most KBOs have been detected by surveys
having limiting magnitudes $m_V \sim 24$ (Millis et al. 2002;
see also the survey statistics compiled in Chiang \& Brown 1999).
For this limiting magnitude, all objects that are inside $r\approx 44\AU$
and that have sizes $s\gtrsim 200\km$ and albedos $\gtrsim 0.04$ would
be detected. These sizes and albedos are comparable to those
estimated by the Minor Planet Center. Moreover, a glance
at Figures \ref{snap32} and \ref{snap21} reveals that indeed,
all but one of the observed Plutino and Twotino candidates have been
discovered at $r < 44\AU$.

Thus, if we assume that $\tau=10^7\yr$,
and take the dotted curve for $r\leq 44\AU$ in Figure \ref{dobjperdeg}
as our guide, then we crudely estimate the bias
in finding Plutinos over Twotinos, averaged over
all longitudes for which Plutinos have been discovered
($330\degr$--$90\degr$, and $150\degr$--$270\degr$),
to be $N_{3:2}/N_{2:1} \sim 2.2$. Then the observed ratio, $F_{obs} \sim 8$,
should be de-biassed down to values closer to
$8/2.2 \sim 3.6$. We emphasize that this is a model-dependent
estimate of $F$ that assumes that $\tau = 10^7\yr$.
Since $(f_{3:2}g_{3:2})/(f_{2:1}g_{3:2})|_{\tau=10^7\yr} \approx 0.40$,
we would estimate the ratio of primeval populations
to be $\eta_0 (a\approx 35\AU)/\eta_0 (a\approx 42\AU) \sim 3.6/0.40 \sim 9$.
This would reflect a steep drop
in mass density over a short distance in the ancient
planetesimal disk. Nonetheless, it would be consonant
with the idea that a ``Kuiper Cliff'' (Chiang \& Brown 1999;
see also Allen, Bernstein, \& Malhotra 2001;
Jewitt, Luu, \& Trujillo 1998; Gladman et al.~1998)
delineates the edge of the classical Kuiper Belt at
$a\approx 48\AU$.

If we assume instead that $\tau = 10^6\yr$ and repeat the same
analysis by using the dotted line for $r \leq 44 \AU$
in Figure \ref{dobjperdegII},
we estimate an intrinsic $F \sim 8/3.0 \sim 2.7$, and
a corresponding ratio of primeval populations
of $\eta_0 (a\approx 35\AU)/\eta_0 (a\approx 42\AU) \sim 2.7/9.6 \sim 0.3$.
Thus, while our estimate that Plutinos intrinsically outnumber Twotinos
by a factor of $\sim$3-to-1 seems robust to changes
in $\tau$,
a dramatic drop in mass density with distance in the primordial
planetesimal disk is by no means an assured conclusion.

The above estimates for $F$ are plagued by other observational biases that
are difficult to quantify. For example, the Minor Planet Center
dataset probably contains more Plutinos relative to Twotinos
than it should because (1) MPC orbit fitting algorithms for objects
discovered a few AU inside Neptune's orbit and having short astrometric
arcs favor Plutino-like orbits as trial solutions
(B.~Marsden 2002, personal communication),
and (2) the astrometric
recovery rate of Plutinos is probably higher than that of
Twotinos because the former objects are, on average, closer and therefore
easier to re-detect by virtue of their brightness and
large apparent proper motion. Both these factors lead
us to conclude that our above estimates for $F$
should be considered upper limits.

Superior estimates for $F$ can be obtained by coupling
our theoretical calculations to the results of surveys having
well-documented discovery statistics and minimal bias in
their algorithms for orbit fitting. The Deep Ecliptic Survey
(DES; Millis et al.~2002), for example, promises to be one
such survey; it employs the more objective method of
Bernstein \& Khushalani (2000) in fitting orbits with short
arcs. We defer analysis of the resonant populations using
their large and homogeneous dataset to future study.

An early but intriguing comparison between theory and observation
lies in the complete absence of observed
Twotino candidates
at longitudes behind that of Neptune (see Figures \ref{snap21}
and \ref{snap21longIIb}). By contrast, our numerical experiments
demonstrate that resonant capture and migration preferentially
fill the 2:1 resonant lobe displaced behind of, rather than ahead of,
Neptune's longitude. Moreover, this asymmetry is only enhanced by faster
rates of migration (Figure \ref{snap21longIIb}).
With only $\sim$5
candidate Twotinos, the probability of finding all of them in the
forward lobe and not the backward lobe is 1/32, under the prior
that a Twotino is as likely to be found in one lobe as the other.
Whether the
actual Kuiper Belt defies the predicted sign of the
asymmetry---in which case the present theory of resonant
capture and migration must be considered either incomplete
or incorrect---only continuing surveys for KBOs will tell.

\section{SUMMARY}
\label{summa}

We have analyzed quantitatively the predictions of the model
of resonant capture and migration for the 2:1 (Twotino)
and 3:2 (Plutino) populations of the Kuiper Belt. We summarize
our main findings as follows:

\begin{enumerate}

\item{The instantaneous spatial distribution of resonant objects
depends not only on the libration centers of their resonant
arguments, $\langle \Phi \rangle$, but also on their
distribution of libration amplitudes, $\Delta \Phi$. For
example, if $\Delta \Phi \gtrsim 1 \rad$ for most Plutinos,
the usual expectation that such objects are most readily
found $\pm 90\degr$ away from Neptune's longitude
would not be valid. The distribution of libration amplitudes
within a given resonance is model-dependent.}

\item{We have numerically evaluated
the capture efficiencies, $f$, of the sweeping
2:1 and 3:2 resonances for three values of
the migration timescale, $\tau(\yr) = 10^7, 10^6,$ and $10^5$.
The timescale is assumed to be one of exponential decay.
We define the capture efficiency to equal the fraction of objects
whose orbits are initially spread uniformly over the complete
path of a given sweeping resonance and that are ultimately captured by it.
This efficiency depends not
only on the probability of capture into the isolated
resonant potential of interest, but also on the probability
of pre-emptive capture into other resonances that lie
exterior to the given resonance, and on the probability
of violent scattering by close encounters with the planets.
The capture efficiency for Twotinos, $f_{2:1}$, decays
from 53\% to 0\% as $\tau$ is reduced from $10^7\yr$
to $10^5\yr$. The capture efficiency for Plutinos, $f_{3:2}$,
increases from 23\% to 78\% as $\tau$ decreases from
$10^7\yr$ to $10^6\yr$; the shorter migration timescale
breeds fewer close encounters and fewer objects are lost to
the competing 2:1 resonance. For $\tau=10^5\yr$, $f_{3:2} \approx 30\%$.}

\item{At least 1 Twotino candidate
having a large eccentricity has been observed. This observation,
interpreted within the confines of our migration model,
implies that $\tau$ cannot be equal to or lower than
$10^5\yr$. This conclusion is subject to the caveat that the migration
model does not fully account for the observations; e.g., the model
fails to generate the large orbital inclinations that are observed
throughout the Kuiper Belt.}

\item{We have simulated the instantaneous spatial distributions
of Twotinos and Plutinos as predicted by the migration model.
If $\tau=10^7\yr$, Twotinos cluster at longitudes displaced
$\pm 75\degr$ away from Neptune's longitude, where the upper
sign corresponds to those objects librating with low
amplitudes, $\Delta \Phi_{2:1} \lesssim 1\rad$, about $\langle \Phi_{2:1}
\rangle \approx \pi/2$, and the lower sign corresponds to objects librating
with
similarly low amplitudes about $\langle \Phi_{2:1} \rangle \approx 3\pi/2$.
Plutinos cluster at longitudes displaced $\pm 90\degr$ away from Neptune's
longitude, and all librate about $\langle \Phi_{3:2} \rangle = \pi$.
Longitude-to-longitude variations in the sky densities of
Plutinos and Twotinos persist even for surveys that are not
volume-limited in their ability to detect resonant objects
of a given size. These variations sharpen as the limiting distance out to which
resonant objects can be detected decreases.}

\item{Over the longitude interval $210\degr$--$240\degr$ ($-90\degr$
to $-60\degr$ of Neptune's longitude), the bias in finding Plutinos
over Twotinos is minimized. If the population of one resonance
is identical to the other in terms of number, sizes, albedos,
and orbital inclinations, then the migration model for $\tau = 10^7\yr$
predicts 0.4 to 1 times as many Twotinos to be found
over this longitude interval than Plutinos, as the limiting
distance out to which objects are detected increases from 40 AU to $\infty$.
If $\tau=10^6\yr$, the bias over this special longitude
range varies from 0.4 to 1.1.
A similar interval exists
between $0\degr$ and $40\degr$ (+$60\degr$--$100\degr$ of Neptune's
longitude) if $\tau=10^7\yr$, but does not exist
if $\tau=10^6\yr$ (see next point).}

\item{The 2:1 resonance fills asymmetrically in the migration
model---more objects are captured into libration about $\langle \Phi_{2:1}
\rangle \approx 3\pi/2$ than about $\langle \Phi_{2:1} \rangle \approx \pi/2$.
The difference betwen populations is $\sim$10\% if $\tau=10^7\yr$, and
increases to $\sim$330\% if $\tau=10^6\yr$. We reserve an
analytic explanation for our numerical discovery to future study.\footnote{We
have performed additional numerical and analytic studies that demonstrate
that this asymmetry can be understood purely within the restricted, circular,
3-body problem where the perturber's orbit expands outward.}
The asymmetry in libration centers translates directly into an asymmetry
in the instantaneous spatial distribution of Twotinos---more Twotinos
are expected to be found at longitudes behind that of Neptune
than in front of it. A differential measurement of the Twotino
density ahead and behind of Neptune would powerfully constrain
the migration history of that planet.}

\item{Measuring the relative populations of Plutinos to Twotinos
is a model-dependent enterprise. Under the assumption that Twotinos
and Plutinos share the same sizes, albedos, and orbital
inclinations, we employ the results in this paper to crudely de-bias
the current tally of $\sim$43 observed Plutinos to $\sim$5 observed
Twotinos to estimate an intrinsic population ratio of $F = 2.7$--3.6.
The range in our estimate reflects an order-of-magnitude difference
in the assumed $\tau$, from $10^7\yr$ to $10^6\yr$.
We suspect that our
estimate is an upper limit because Plutinos probably enjoy a greater
frequency of astrometric recovery than Twotinos, and because
orbit fitting algorithms employed by the Minor Planet Center, whose
dataset we use, favor Plutino-like trajectories.
Improved estimates for $F$ can be obtained by coupling our calculation to KBO
surveys having well-documented discovery statistics.
While it tentatively appears that Plutinos might intrinsically
outnumber Twotinos by a factor not exceeding $\sim$3,
this conclusion can but does
not necessarily imply that the surface density of the primordial
planetesimal disk dropped dramatically with distance in the vicinity
of $\sim$42 AU.}

\end{enumerate}

\acknowledgements
This work was supported by a National Science Foundation
Planetary Astronomy Grant AST-0205892, two Faculty Research Grants
awarded by the University of California at Berkeley,
and a UC Berkeley Letters \& Science Undergraduate Research Travel Grant
awarded to ABJ. We thank Marc Buie, Jim Elliot, Susan Kern, Renu Malhotra,
Bob Millis, David
Trilling, and Larry Wasserman for encouraging discussions,
and an anonymous referee for providing a critical and insightful
report.

\newpage
\begin{deluxetable}{lcccccccc}
\tabletypesize{\scriptsize}
\tablewidth{0pt}
\tablecaption{Summary of Simulations}
\tablehead{
\colhead{Label} &
\colhead{$\tau (\YR)$} & \colhead{Resonance} &
\colhead{$f$(\%)\tablenotemark{a}} &
\colhead{$g$(\%)\tablenotemark{b}} &
\colhead{$\langle \Phi \rangle \approx \pi$ (\%)\tablenotemark{c}} &
\colhead{$\langle \Phi \rangle \approx 3\pi/2$ (\%)\tablenotemark{c}} &
\colhead{$\langle \Phi \rangle \approx \pi/2$ (\%)\tablenotemark{c}} &
\colhead{Figures}
}
\startdata
Ia & $10^7$ & 3:2 & 23 & 46 & 100 & 0 & 0 &
3,5,6,10--12,\ref{snap32},\ref{objperdeg},\ref{dobjperdeg} \\
Ib & $10^7$ & 2:1 & 53 & 50 & 16 & 44 & 40 & 4,5,7--11,\ref{w21},15--17 \\
IIa & $10^6$ & 3:2 & 78 & 92 & 100 & 0 & 0 &
\ref{histo32allin1},\ref{objperdegII},\ref{dobjperdegII} \\
IIb & $10^6$ & 2:1 & 15 & 50 & 13 & 67 & 20 &
\ref{lib21IIb},\ref{snap21longIIb}--\ref{dobjperdegII} \\
IIIa & $10^5$ & 3:2 & 30 & 97 & 100 & 0 & 0 & \ref{histo32allin1}\\
IIIb & $10^5$ & 2:1 & 0 & $\ldots$ & $\ldots$ & $\ldots$ & $\ldots$ & $\ldots$
\\
\enddata
\tablenotetext{a}{Efficiency of capture into a given resonance.}
\tablenotetext{b}{Efficiency of retainment of captured objects
by a given resonance
over the age of the solar system. For the 2:1 resonance, this is
assumed to be 50\% (Malhotra 2002, personal communication).
For the 3:2 resonance, we take the
retained population to comprise those captured objects that either have
$\Delta \Phi_{3:2} < 110\degr$ or that inhabit Kozai-type
resonances for which $\langle \omega \rangle = \pm 90\degr$
(compare with Levison \& Stern 1995).}
\tablenotetext{c}{Percentage of objects in the resonance that librate
approximately about
the mean value indicated.}
\label{capefftab}
\end{deluxetable}

\end{document}